\documentclass[10pt,journal,compsoc]{IEEEtran}

\ifCLASSOPTIONcompsoc
  \usepackage[nocompress]{cite}
\else
  \usepackage{cite}
\fi
\usepackage{amsmath,amsfonts}
\usepackage{algorithmic}
\usepackage{algorithm}
\usepackage{array}
\usepackage[caption=false,font=normalsize,labelfont=sf,textfont=sf]{subfig}
\usepackage{textcomp}
\usepackage{stfloats}
\usepackage{url}
\usepackage{verbatim}
\usepackage{graphicx}
\usepackage{cite}
\usepackage{bm}
\usepackage{amssymb}
\usepackage{booktabs}
\usepackage{overpic}
\usepackage[breaklinks=true]{hyperref}
\hypersetup{
	pdfauthor = {},
	pdftitle = {},
	pdfsubject = {},
	pdfkeywords = {},
	colorlinks=true,
	linkcolor= blue,
	citecolor= blue,
	pageanchor=true,
	urlcolor = blue,
	plainpages = false,
	linktocpage
}

\usepackage{amssymb} 
\usepackage{microtype} 
\usepackage{enumitem}
\usepackage{booktabs}
\usepackage{multirow}
\usepackage{makecell} 
\usepackage{booktabs}
\usepackage[table]{xcolor}

\newcommand{\paraheading}[1]{\vspace{0.5em}\noindent\textbf{#1}.~}

\newcommand{\frontparaheading}[1]{\noindent\textbf{#1}.~}

\ifCLASSINFOpdf
 
\else

\fi

\begin{document}
\title{Computational Caustic Design for Surface Light Source}

\author{Sizhuo~Zhou,
        Yuou~Sun,
        Bailin~Deng,~\IEEEmembership{Member,~IEEE,}
        and~Juyong~Zhang$^\dagger$,~\IEEEmembership{Member,~IEEE}
\IEEEcompsocitemizethanks{\IEEEcompsocthanksitem  S. Zhou is with the School
of Mathematical Sciences, University of Science and Technology of China and Shanghai Innovation Institute.
}
\IEEEcompsocitemizethanks{\IEEEcompsocthanksitem  Y. Sun, and J. Zhang are with the School
of Mathematical Sciences, University of Science and Technology of China.
}
\IEEEcompsocitemizethanks{\IEEEcompsocthanksitem  B. Deng is with the School of     
        Computer Science and Informatics, Cardiff University.}

	\thanks{$^\dagger$Corresponding author. Email: \texttt{juyong@ustc.edu.cn}.}
}

\markboth{IEEE TRANSACTIONS ON VISUALIZATION AND COMPUTER GRAPHICS}%
{Shell \MakeLowercase{\textit{et al.}}: Bare Demo of IEEEtran.cls for Computer Society Journals}

\IEEEtitleabstractindextext{%
\begin{abstract}
Designing freeform surfaces to control light based on real-world illumination patterns is challenging, as existing caustic lens designs often assume oversimplified point or parallel light sources. We propose representing surface light sources using an optimized set of point sources, whose parameters are fitted to the real light source's illumination using a novel differentiable rendering framework. Our physically-based rendering approach simulates light transmission using flux, without requiring prior knowledge of the light source's intensity distribution. To efficiently explore the light source parameter space during optimization, we apply a contraction mapping that converts the constrained problem into an unconstrained one. Using the optimized light source model, we then design the freeform lens shape considering flux consistency and normal integrability. Simulations and physical experiments show our method more accurately represents real surface light sources compared to point-source approximations, yielding caustic lenses that produce images closely matching the target light distributions.
  
\end{abstract}

}

\maketitle

\IEEEdisplaynontitleabstractindextext

\begin{figure*}[ht]
  \centering
  \includegraphics[width=\linewidth]{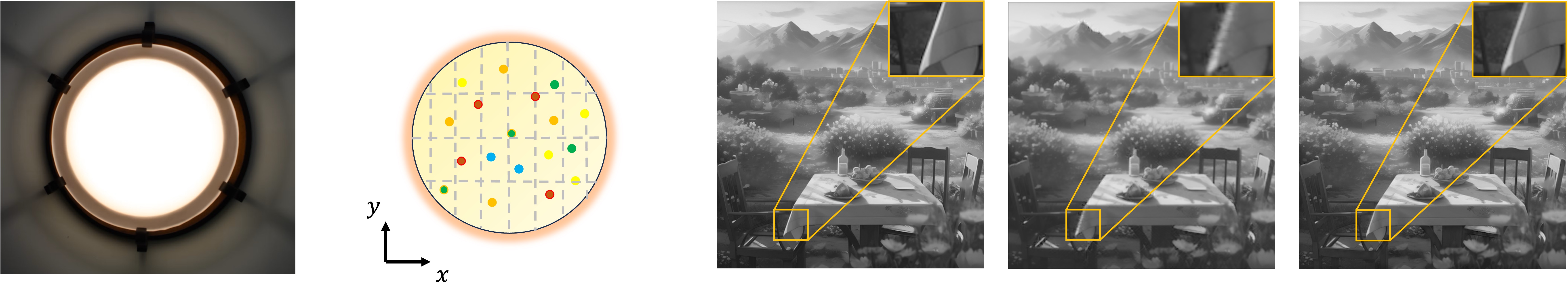}\\
  \vspace{-1em}
  \begin{picture}(0,0)
    \put(-237,-4){\footnotesize Realistic Surface}
    \put(-230,-14){\footnotesize Light Source}
    \put(-123,-10){\footnotesize Light Source Model}
    \put(-2,-10){\footnotesize Target Image}
    \put(87,-4){\footnotesize Point Light Source}
    \put(108,-14){\footnotesize Based}
    \put(196,-10){\footnotesize Our Method}
  \end{picture}
  \vspace{1em}
  \caption{\label{teaser} 
    The rendered results of the optimized freeform surface lens under the illumination of a surface light source according to the target image. The above lenses were obtained through the caustic design based on a point light source and an optimized surface light source model, respectively. The light pattern produced by the caustic lens based on our method closely resembles the target image. Some regions are enlarged for better inspection of details.}
  \end{figure*}
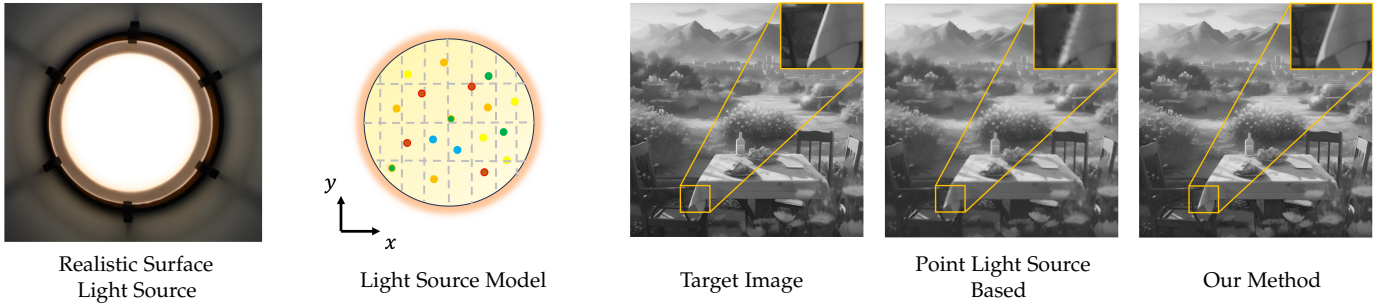
  
\IEEEpeerreviewmaketitle

\IEEEraisesectionheading{\section{Introduction}\label{sec:introduction}}

\IEEEPARstart{T}{he} interaction between light and objects can produce various captivating illumination effects, such as caustic phenomena that occur when light undergoes refraction or reflection. 
Precisely controlling this caustic process by designing freeform surfaces has important applications in computer graphics~\cite{schwartzburg2014high, meyron2018light} and geometric optics~\cite{budhu2019novel, kass2021}. 
Additionally, accurately controlling the caustic process to achieve desired flux distributions has broad applications in fields such as art design, interior lighting, and industrial manufacturing.

In the field of non-imaging optics, the shape of freeform surface lenses is typically designed to control various properties of light in order to create the desired illumination effects. 
Freeform surfaces here refer to optical surfaces without linear or rotational symmetry~\cite{reimers2017freeform}. 
They offer a high degree of freedom, thereby enabling precise control over light beams.

Existing methods in computer graphics typically reconstruct the final freeform surface by calculating certain shape characteristics of the required lens, such as its normals~\cite{schwartzburg2014high} or visibility diagram~\cite{meyron2018light}. 
However, these methods almost always assume Lambertian point light sources or parallel light~\cite{Shen21, schwartzburg2014high, meyron2018light}, which are inaccurate for real-world scenarios as ideal point or parallel light sources do not exist. Instead, real light sources often have a finite area. This discrepancy affects the accuracy of caustic design, causing the designed lenses to produce blurred images on the receiving plane.

The root cause of this issue is the failure to accurately simulate the emission pattern of LED surface light sources during the design process of freeform caustic lenses. When using a point light source or parallel light for illumination simulation, each point in space receives light from only one direction. In contrast, with a surface light source, each position in space receives light from multiple directions emitted by the extended source. Especially when the light source is close to the object or is itself large, using a Lambertian point light source to simulate actual illumination will significantly compromise the accuracy of caustic design.

Additionally, existing methods for simulating surface light source illumination typically sample rays based on the actual light intensity distribution on the surface~\cite{xu2025hybrid, zli2020, sun2021end}. 
This deviates from true physical behavior and incurs high computational costs. More critically, the exact light intensity distribution of an LED surface light source is often unknown, making it impossible for ray-sampling-based methods to accurately simulate illumination~\cite{raytracing1984}.

To overcome these issues, we propose a novel freeform caustic lens design scheme based on surface light sources. Our method is divided into two parts: 1)~modeling and optimizing an LED surface light source representation, and 2)~optimizing the lens shape under the optimized surface light source model. We first propose a novel surface light source representation strategy and construct a fully differentiable rendering framework based on the light path control characteristics of freeform surfaces. This rendering framework accurately simulates the propagation and refraction behavior of light and can be used for both optimization of our light source model and optimization of the caustic lens shape.

Specifically, we first select a set of lenses with known shapes represented by triangle meshes. We then use the differences in brightness distribution between the rendered patterns, produced by the interaction of the light source model with these lenses, and the corresponding reference patterns to drive the optimization of the model parameters. This approach enables the model to continuously improve its ability to accurately represent the illumination characteristics of the surface light source. Afterwards, using the optimized area light source model, we constructed an optimization scheme for the shape of the freeform surface lens.

It is worth noting that optimizing the parameters of the surface light source model with the help of caustic lenses with known shapes is a crucial aspect of our design.
In the optimization process of the model, the introduction of caustic lenses can amplify the differences in the propagation directions of various light rays, clearly highlighting the illumination characteristics of the surface light source.
At the same time, using caustic lenses can form meaningful patterns on the receiving plane, making it easier for us to evaluate image details.
To address the issue of imprecision in ray sampling strategies, we introduced the concept of light flux. By calculating the light flux within a region, we achieve an exact simulation of the light propagation process.
This light energy transfer approach is more accurate in a physical sense and does not require prior knowledge of the intensity distribution on the light source surface when combined with our light source model.
Furthermore, considering that the inherent size of the surface light source introduces boundary constraints, we designed a contraction mapping~\cite{barron2022mipnerf360} to transform the constrained optimization problem into an unconstrained one, which can be efficiently solved by using the L-BFGS algorithm~\cite{Liu1989OnTL}.
To further reduce the time cost of the rendering process, we incorporated a parallel design into the optimization process.

This paper addresses a fundamental limitation in computational caustic design: the rendering blur caused by idealized point or parallel light source assumptions. By introducing a physically-based, optimizable model for surface light sources, our method significantly improves the accuracy of caustic generation (see Figure~\ref{teaser}). 
This advance is critical for a variety of practical applications where precision is important, including artistic design~\cite{kiser2013architectural}, industrial lighting~\cite{zhu2013optical, kim2013design}, solar energy collection~\cite{wallhead2012design, xie2011concentrated, leutz2012nonimaging}, and the design of advanced optical instruments and lenses for medical, biochemical~\cite{goure2013optics, rovati1996design}, and astronomical applications~\cite{bland2009astrophotonics}.

In summary, our contributions include:
\begin{itemize}[leftmargin=*]
    \item A concise and effective representation model for surface light sources, which can reflect the lighting characteristics of real light sources;
    \item A differentiable rendering framework and algorithm to optimize model parameters based on this rendering framework;
    \item A freeform surface optimization method that applies an optimized surface light source model, which results in a rendered image closer to the target image compared to a point light source-based caustic design.
\end{itemize}

\begin{figure*}[tbp]
  \centering
  \mbox{} \hfill
  \hfill
  \includegraphics[width=1\linewidth]{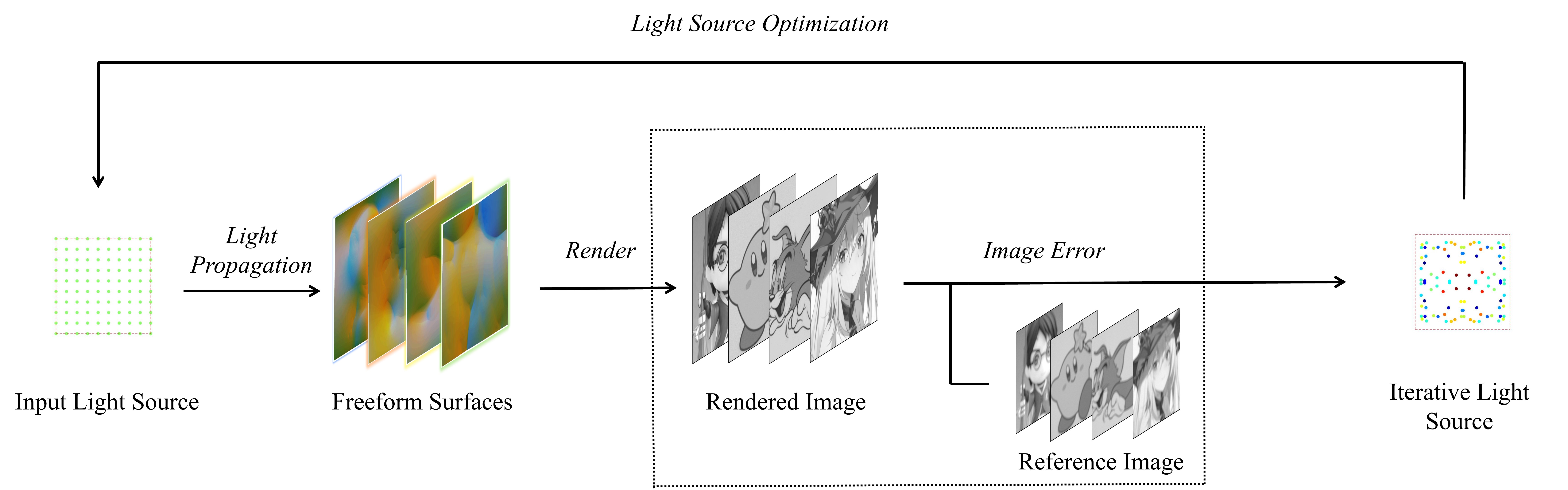}
  \hfill \mbox{}
  \vspace{-1em}
  \caption{\label{pipeline1}
    Light source model optimization process: optimizing the surface light source model based on the difference between the rendering result and the reference image.}
\end{figure*}

\section{Related Work}

\frontparaheading{Freeform caustic surface design}
Freeform surfaces are special optical devices commonly used to achieve beam control.
Specifically, by designing a high degree of freedom optical surface (usually a lens), the light emitted by the light source is redirected to produce a specified irradiance distribution~\cite{winston2005nonimaging}.
Compared with traditional optical surfaces (including spherical, non-spherical, and cylindrical surfaces), due to the huge degree of freedom of freeform surfaces, the performance of optical systems can reach new heights in fields such as imaging~\cite{Rolland:21, Mao:18, Cheng:21}, illumination~\cite{Rengmao2018Design, Li18}, and laser beam shaping~\cite{Feng2021Iterative}.
In fact, many real-life scenarios require specialized optical systems, such as automobile headlights~\cite{zhu2013optical} and projectors~\cite{damberg2016high, zhao2013integral}.
Thus, it can be seen that freeform surfaces play an important role in achieving specific control of light beams~\cite{Duerr15, Li18}.

Another type of work was first proposed by Finckh et al.~\cite{finckh2010geometry}. 
They used freeform surfaces to refract or reflect light to form the desired caustic image.
This work uses a stochastic approximation optimization algorithm to determine the surface geometry from the caustics.
This requires a large amount of computation and cannot achieve accurate control of areas with weak light intensity.
Subsequent work starts from the principle of local energy conservation and formulates it as a problem of solving a class of Monge-Ampère equations~\cite{zhu2013optical, ZHANG2014297, Chang_2016}.
These methods assume that the lens shape is a continuous and smooth surface, thereby reducing artifacts to achieve higher-quality images.
Some other works solve the shape of the freeform surface by means of the integral normal field~\cite{kiser2013architectural} or formulate it as a Poisson reconstruction problem~\cite{yue2014poisson}.
However, the freeform surfaces obtained based on these methods cannot produce high-contrast images.
Schwartzburg et al.~\cite{schwartzburg2014high} solved this problem. 
They obtained the lens surface by calculating the normal of the lens surface through optimal transport, which can produce a pure black area on the receiving plane.
Meyron et al.~\cite{meyron2018light} proposed an algorithm to solve the optimal transport problem using visibility maps, further reducing the loss of light energy.
Marc et al.~\cite{Kassubeck2021NSfCRA} combined caustic design with novel neural denoising components and a learned gradient descent optimizer, improving the stability and quality of the iterative reconstruction process.

\paraheading{Differentiable rendering}
Differentiable rendering is included in the field of inverse rendering, which estimates physical parameters from imaging data using a differentiable renderer. 
These rendering methods describe the image formation process in a differentiable way and can effectively calculate the gradients of relevant free parameters.
Therefore, they can be used in parameter search strategies based on gradient descent.
In caustic design, differentiable rendering is used to obtain images generated by the joint action of illumination sources and free lenses, as well as the gradients of relevant physical parameters.
However, traditional rendering methods usually include non-differentiable components. 
Rasterization~\cite{pineda1988parallel} is a typical example.
Therefore, it is not easy to directly optimize the physical properties in the rendering scene based on the simulation results.
In recent years, many works have been studying how to make the rendering process differentiable.
Liu et al.~\cite{liu2019soft} proposed soft rasterization, which uses probability distributions and aggregation functions to make the rendering pipeline differentiable. 
Other works~\cite{Bangaru2020, Li2012Diff, Zhang2020Path, Zhang2021PSDR} solve the problem of computing gradients with respect to the scene geometry, which is not differentiable due to the discontinuity of the mesh.
After that, some differentiable renderers have emerged, such as Mitsuba2~\cite{Loubet2019, NimierDavid2020, NimierDavidVicini2019}, Mitsuba3~\cite{Mitsuba3, Jakob2020DrJit}, and Path Replay Backpropagation~\cite{Vicini2021}.
The ray tracing-based approach in Mitsuba3 can be applied to caustic design. 
However, since this renderer is not specifically tailored for caustic design, there remains considerable room for improvement in its rendering results.
Wang et al.~\cite{wang2022differentiable} proposed a strategy based on ray tracing and deep lens systems, which can be applied to a variety of inverse design problems.
However, certain limitations still remain regarding its performance and applicability.

In addition to path tracing algorithms, photon mapping is also a classic global illumination algorithm. 
Originally proposed by Jensen in 1996~\cite{jensen1996global}, it efficiently simulates complex lighting effects such as caustics by tracing photon paths and constructing a photon map.
To optimize memory usage, Hachisuka et al.~\cite{Hachisuka2008} introduced Progressive Photon Mapping (PPM), which performs photon tracing in multiple iterations.
Several other studies have also introduced various improvements to the photon mapping algorithm.
Stochastic Progressive Photon Mapping (SPPM)~\cite{Hachisuka2009} enabled the support of effects such as glossy reflections. 
Georgiev et al.~\cite{Georgiev2012} and Hachisuka et al.~\cite{Hachisuka2012} reformulated photon mapping as a vertex merging technique compatible with path integration, which laid the foundation for differentiable photon mapping.
Xing et al.~\cite{xing2024differentiable} leveraged Extended Path Space Manifolds (EPSMs) to achieve differentiability of path vertex positions and color contributions with respect to the scene parameters. 
However, the rendered results still exhibit significant noise and continue to require light path sampling.
To achieve a physically more accurate caustic design, we use a differentiable framework based on flux transport.

\paraheading{Optimal Transport in Caustic Design}
The problem of caustic lens design is essentially an issue of adjusting the light field distribution.
Finding the correspondence between two density distributions is an important problem in mathematics, and optimal transport (OT) is an effective tool for solving this problem.
It finds an optimal way to transfer one mass distribution into another, minimizing the total transportation cost.
This problem was first proposed by the French mathematician Gaspard Monge in 1781~\cite{monge1781memoire} and was later extended by Kantorovich to allow for the splitting of mass~\cite{Kantorovich2006Translocation}.
The classic optimal transport problem includes both discrete and continuous forms. Subsequently, an important semi-discrete form emerged, which seeks an optimal mapping that converts a discrete distribution into a continuous one~\cite{levy2015numerical, merigot2011multiscale}.
\cite{schwartzburg2014high, Oliker:07, Oliker2011Designing, Feng2021Iterative} all adopt the optimal transport strategy.
However, this strategy is only suitable for caustics based on point light sources or parallel light.
Taking the triangle mesh representation of a lens as an example, when the input light source is a surface light source, each triangular area on the lens surface no longer controls only a single light ray. 
Our approach solves the above issue.


\section{Method}
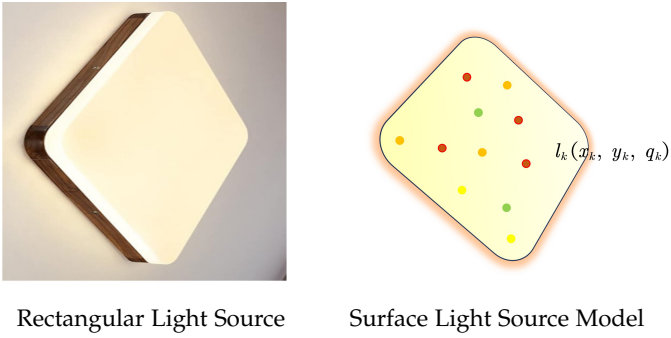
\begin{figure}[t]
  \centering
  \includegraphics[width=1.0\linewidth]{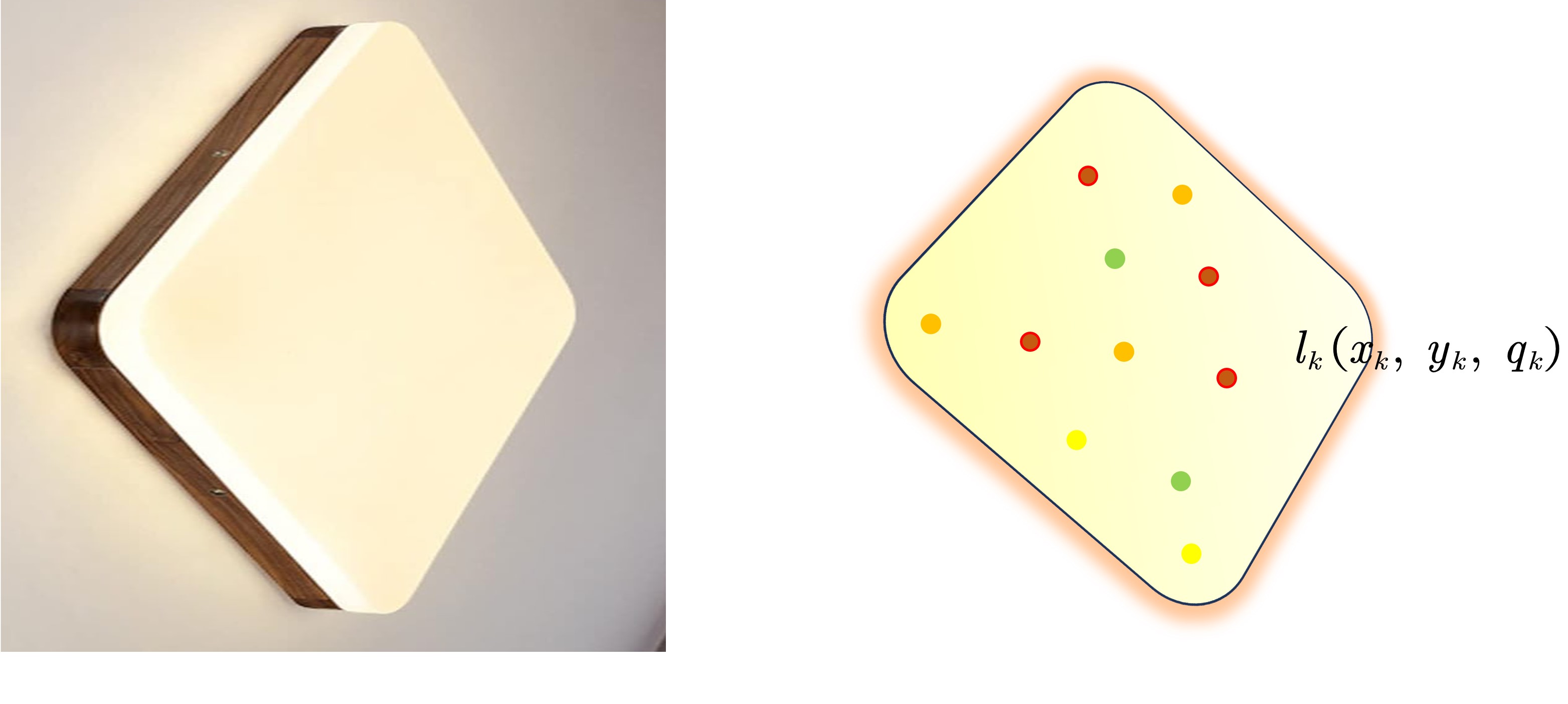}\\
  \vspace{-0.5em}
  \begin{picture}(0,0)
    \put(-121,-2){\small Rectangular Light Source}
    \put(5,-2){\small Surface Light Source Model}
  \end{picture}
  \caption{\label{surface_ls}
           The surface light source model. Each point light source in this model contains three attributes: $x$-coordinate, $y$-coordinate, and its illumination intensity $q_k$.}
\end{figure}

\begin{figure}[t]
  \centering
  \includegraphics[width=0.98\linewidth]{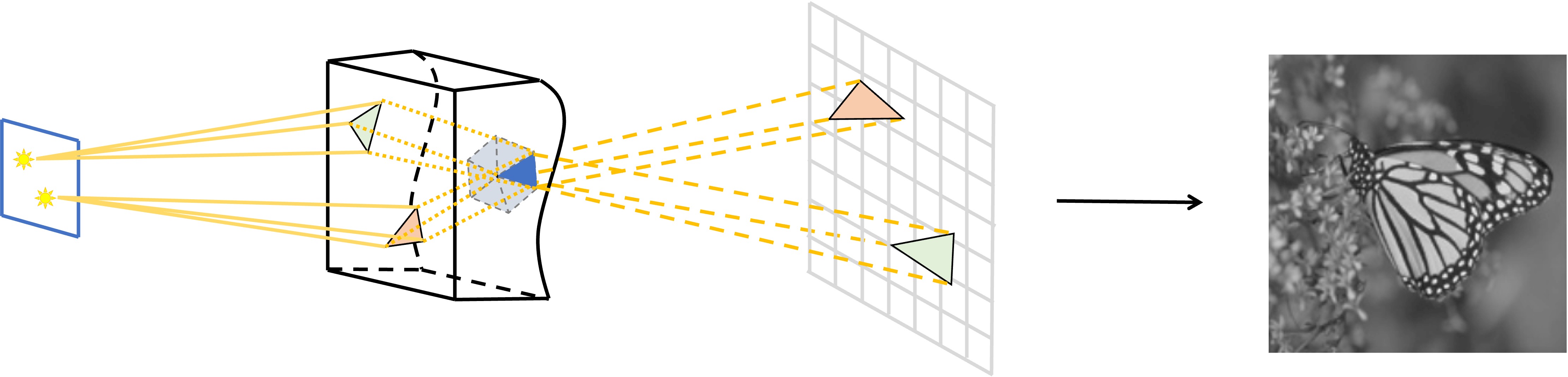}\\
  \vspace{-1.6em}
    \begin{picture}(0,0)
    \put(-125,-12){\footnotesize Light Source}
    \put(-108,-22){\footnotesize Model}
    \put(-59,-16){\footnotesize Lens}
    \put(-4,-16){\footnotesize Receiving Plane}
    \put(75,-16){\footnotesize Rendered Image}
  \end{picture}
  \vspace{1.8em}
  \caption{\label{rendering_model}
           Our rendering model. It consists of three components: an input light source, a lens, and a receiving plane. The light undergoes two refractions through the front and back surfaces of the lens before forming an image on the receiving plane.}
\end{figure}
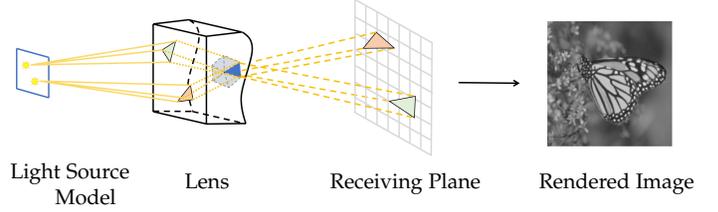

We aim to develop a design method for the shape of caustic lenses based on surface light sources. Given a surface light source and a target image, our computational caustic design approach consists of two key steps:
\begin{enumerate}[leftmargin=*]
    \item First, we approximate the surface light source with a small number of point light sources (see Figure~\ref{surface_ls}), and optimize their positions and intensities by minimizing the differences between their caustic images on a set of reference lens shapes. These point light sources constitute our light source model.
    \item Afterwards, we utilize the optimized light source model and rendering models (see Figure~\ref{rendering_model}) to obtain the rendered images, while simultaneously optimizing the lens shape based on the differences between these images and the target images.
\end{enumerate}

Essentially, the rendered images obtained based on our light source model are the superposition of caustic images formed by individual point light sources in terms of flux. It is worth noting that both steps use the same differentiable method to model the rendering process. The main differences between the two steps lie in the variables being optimized and some design details, such as the loss terms. In the following, we present the details of our rendering model.

\subsection{Surface light source representation model}
For simplicity, we consider surface light sources with planar shapes in this paper, i.e., the light-emitting surface can be represented as a 2D region $D$ 
within the plane $z = 0$. 
To optimize a lens shape according to a target image and a surface light source, we need to derive a rendering model that determines the caustic image for any given lens shape. However, since the planar light source contains a continuous light-emitting region, direct computation of the caustic image would require integrating the caustics resulting from each point in the region, which is complicated and time-consuming. 

To simplify the rendering model, we approximate the planar light source using $N$ ideal point light sources $\{l^k \mid k=1,2,\ldots,N\}$. 
Each point light source $l^k$ has three attributes: the local $x$- and $y$-coordinates $(x_k, y_k)$ in the light source plane, and the light intensity weight $q_k$. 
We can easily determine the resulting caustic image of these $N$ point light sources by superposition of the caustics from each individual point light source.

To approximate the surface light source, we simultaneously optimize the attributes  $\{(x_k, y_k, q_k)\}$ of all point light sources such that the caustic image is as close to the ground-truth caustics of the surface light source as possible. This optimization process ensures that our light source model faithfully represents the illumination characteristics of the original surface light source.


\subsection{Luminous flux}
After modeling the surface light source, we introduce the concept of luminous flux to render the illumination intensity of each point light source onto every pixel of the resulting image. 
In our rendering model, for a certain area on the front surface of the lens, its luminous flux is determined by the solid angle generated by the illumination of point light sources. 
The luminous flux on the back surface of the lens is determined according to the corresponding relationship with the area on the front surface, which we call the source flux.

\paraheading{Source flux}
We assume that the front surface of each lens is a plane positioned parallel to the light source plane, while the back surface is a freeform surface represented by a triangle mesh. 
We denote the domain of the front surface as $U \subset \mathbb{R}^2$. The back surface can then be represented as a height field over $U$. 

\begin{figure}[t]
  \centering
  \includegraphics[width=1.0\linewidth]{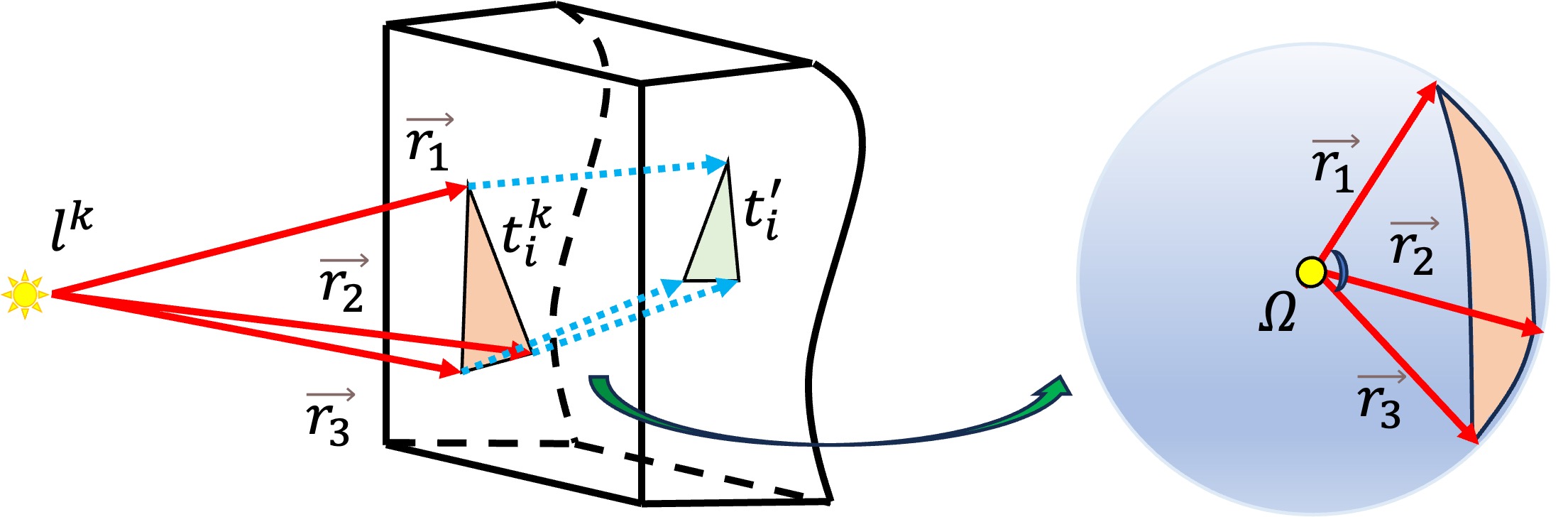}
  \caption{\label{solid_angle}
           Schematic diagram of solid angle calculation: $\vec{r_1}$,\ $\vec{r_2}$ and $\vec{r_3}$ are the unit vectors connecting $l^k$ and the three vertices of $t_i^k$.}
\end{figure}

In this paper, we assume the front and back surfaces of the lens are regulated with the same connectivity and with one-to-one correspondence between their vertices and edges. For a single point light source $l^k$, to determine the flux passing through a triangle $t'_i$ on the back surface, we use Snell's law to compute its corresponding triangle $t_i^k$ on the front surface (see Section~\ref{diff_rendering} for more details). 
Assuming no energy loss during the refraction process, we have 
\begin{equation}
\label{source_flux}
\Phi^k(t'_i) = \Phi^k(t_i^k), 
\end{equation}
where $\Phi^k(I)$ represents the flux passing through the region $I$ given by $l^k$.
Note that the flux $\Phi^k(t_i^k)$ is directly proportional to the solid angle $\Omega_i^k$ of the triangle $t_i^k$ with respect to the point light source $l^k$,  which is the area of the spherical triangle formed by the three unit vectors connecting $l^k$ and the three vertices of $t_i^k$ (see Figure~\ref{solid_angle}) and can be computed as 
\begin{equation}
\Omega_i^k = 2 \arctan \left(\frac{\vec{r_1} \cdot (\vec{r_2} \times \vec{r_3})}{1 + \vec{r_1} \cdot \vec{r_2} + \vec{r_1} \cdot \vec{r_3} + \vec{r_2} \cdot \vec{r_3}} \right),
\end{equation}
After normalization, we obtain the flux $\Phi^k(t'_i)$ passing through triangle $t'_i$ given by the point light source $l^k$. 

\subsection{Differentiable rendering}

\label{diff_rendering}
In order to solve the corresponding relationship of triangular areas on the front and back surfaces of the lens and on the receiving plane, we need to model the process of light propagation and rendering.
Given a freeform surface represented by a triangle mesh and an input light source, our objective is to develop a differentiable algorithm to compute the flux distribution on the receiving plane, and adjust the distribution to closely match the reference image.

\begin{figure}[t]
  \centering
  \includegraphics[width=1.0\linewidth]{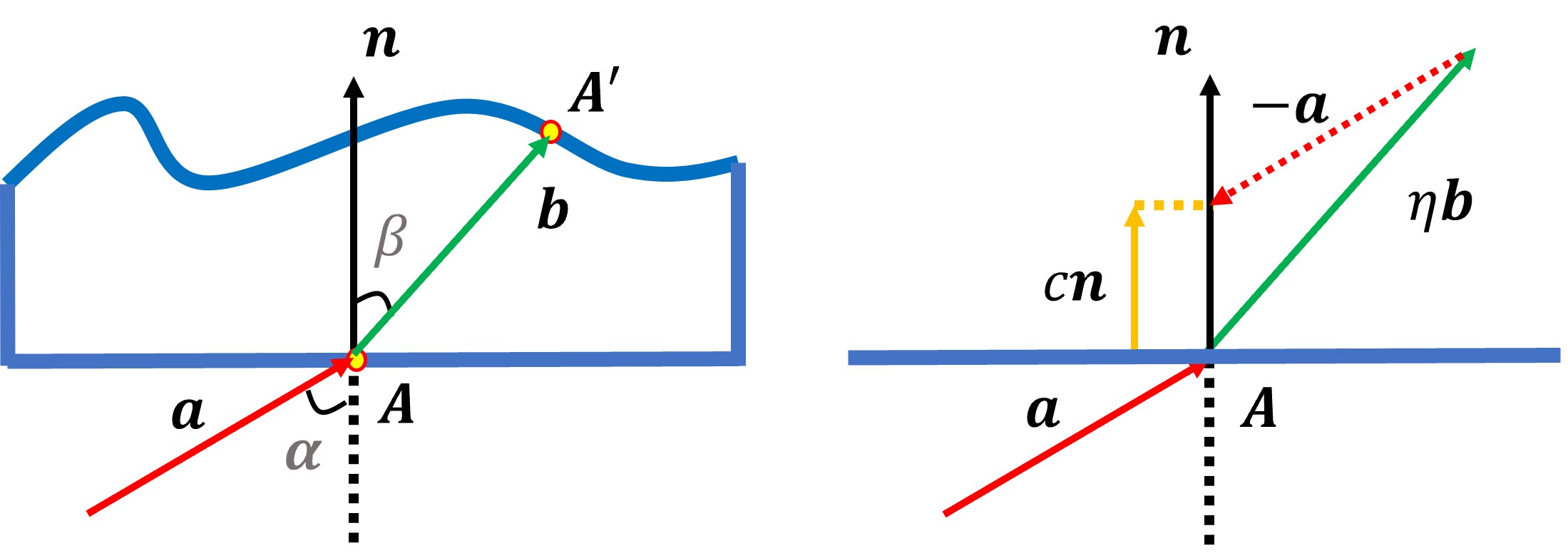}\\
  \vspace{-0.5em}
  \begin{picture}(0,0)
    \put(-100,-4){\small Refractive process}
    \put(42,-4){\small Law of refraction}
  \end{picture}
  \vspace{0.3em}
  \caption{\label{snell_law}
           From left to right: the refraction process of light on the front surface of the lens; according to Snell's law, the relationship between the incident light $\mathbf{a}$, refracted light $\mathbf{b}$ and the surface normal $\mathbf{n}$.}
\end{figure}
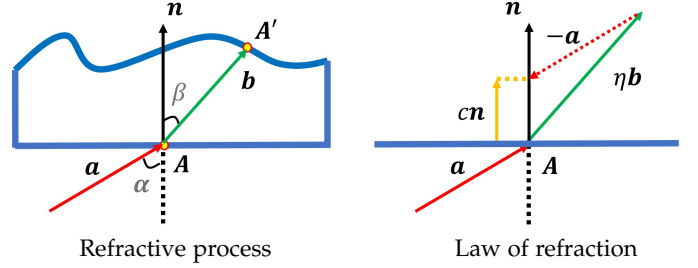

\paraheading{Reverse solution of incident triangle}
Assuming the position of the light source $l^k$ has been given (without loss of generality, translate it to the origin $O$), we first need to determine the position of triangle $t_i^k$ on the front surface. In other words, for each vertex $A'(x_{A'}, y_{A'}, z_{A'})$ on the back surface, we want to find the refraction point $A$, that is, the intersection point of the light ray passing through $A'$ and the front surface. Assume that the front surface lies in the plane $z = z_0$ and has a normal $\mathbf{n}=(0, 0, 1)$, then the coordinate of $A$ can be expressed as $(kx_{A'}, ky_{A'}, z_0)$, where $k$ is an unknown constant. According to Snell's Law  (see Figure~\ref{snell_law}), we can deduce that 
\begin{equation}
(\eta \mathbf{b} - \mathbf{a}) \propto \mathbf{n},
\end{equation}
where $\eta$ is the refractive index of the medium, $\mathbf{b}$ and $\mathbf{a}$ are unit vectors of $\overrightarrow{AA'}$ and $\overrightarrow{OA}$, respectively. Using this relationship, we can further derive the following quartic equation about $k$:
\begin{equation}
\begin{aligned}
\label{quartic}
(\eta^2 - 1)(x_{A'}^2 + y_{A'}^2) \cdot k^4 - 2(\eta^2 - 1)(x_{A'}^2 + y_{A'}^2) \cdot k^3 \\ 
+ [(\eta^2 - 1)(x_{A'}^2 + y_{A'}^2) + \eta^2 \cdot z_0^2 - (z_{A'} - z_0)^2] \cdot k^2 \\ 
- 2 \eta^2 z_0^2 \cdot k + \eta^2 z_0^2 = 0.
\end{aligned}
\end{equation}

It is easy to prove the existence of solutions to the quartic equation within the interval $[0,1]$. 
Therefore, we can solve the equation to obtain the specific coordinates of the intersection point $A$ on the front surface. 
If the position of the point light source is not at the origin, a coordinate transformation is required to convert it into the same form. Note that to ensure the model remains differentiable, algorithms for solving approximate solutions of high-order equations cannot be used here. 

Recall that for any triangular region $t_i'$ on the back surface of the lens, its source flux $ \Phi^k(t_i')$ is equal to the source flux $\Phi^k(t_i)$ of the corresponding triangle $t_i$ on the front surface (see Equation~\eqref{source_flux}). 
Therefore, the flux of all triangles on the back surface can be determined by calculating the flux of the corresponding triangles on the front surface. 
Next, we will establish the flux relationship between the triangular area on the back surface of the lens and each pixel area on the receiving plane.

\paraheading{Target flux}
The light passing through the triangle $t_i'$ on the surface will undergo a second refraction, and the exit direction $\mathbf{b}'$ can be calculated by the following equation derived from Snell's law: 
\begin{equation}
\label{refraction_rear}
\mathbf{b}' = \mathbf{n}' \sqrt{1 + \eta^2 ((\mathbf{a}' \cdot \mathbf{n}')^2 - 1)} + \eta (\mathbf{a}' - (\mathbf{a}' \cdot \mathbf{n}') \mathbf{n}'),
\end{equation}
where $\mathbf{a}'$ and $ \mathbf{n}'$ are the unit vector of incident light and the normal of triangle $t_i'$, respectively. 
Since the back surface is represented as a triangle mesh, each point inside triangle $t_i'$ has the same surface normal:
\begin{equation}
\mathbf{n}'_i = \frac{ (\mathbf{v}_i^2 - \mathbf{v}_i^1) \times (\mathbf{v}_i^3 - \mathbf{v}_i^1) }{ \| (\mathbf{v}_i^2 - \mathbf{v}_i^1) \times (\mathbf{v}_i^3 - \mathbf{v}_i^1) \| },
\end{equation}
where $\mathbf{v}_i^1, \mathbf{v}_i^2, \mathbf{v}_i^3$ are the vertices of $t'_i$ in the positive orientation. 
The light passing through $t_i'$ will intersect with the receiving plane, forming a projection triangle area $t_i''$. 
Each vertex of triangle $t_i''$ is located on the outgoing ray $\mathbf{b}_i'$ corresponding to the vertex of triangle $t_i'$ on the back surface of the lens. 
Since there is no energy loss during the refraction process, the following holds: 
\begin{equation}
\Phi^k(t''_i) = \Phi^k(t'_i).
\end{equation}

\begin{figure}[t]
  \centering
  \includegraphics[width=1.0\linewidth]{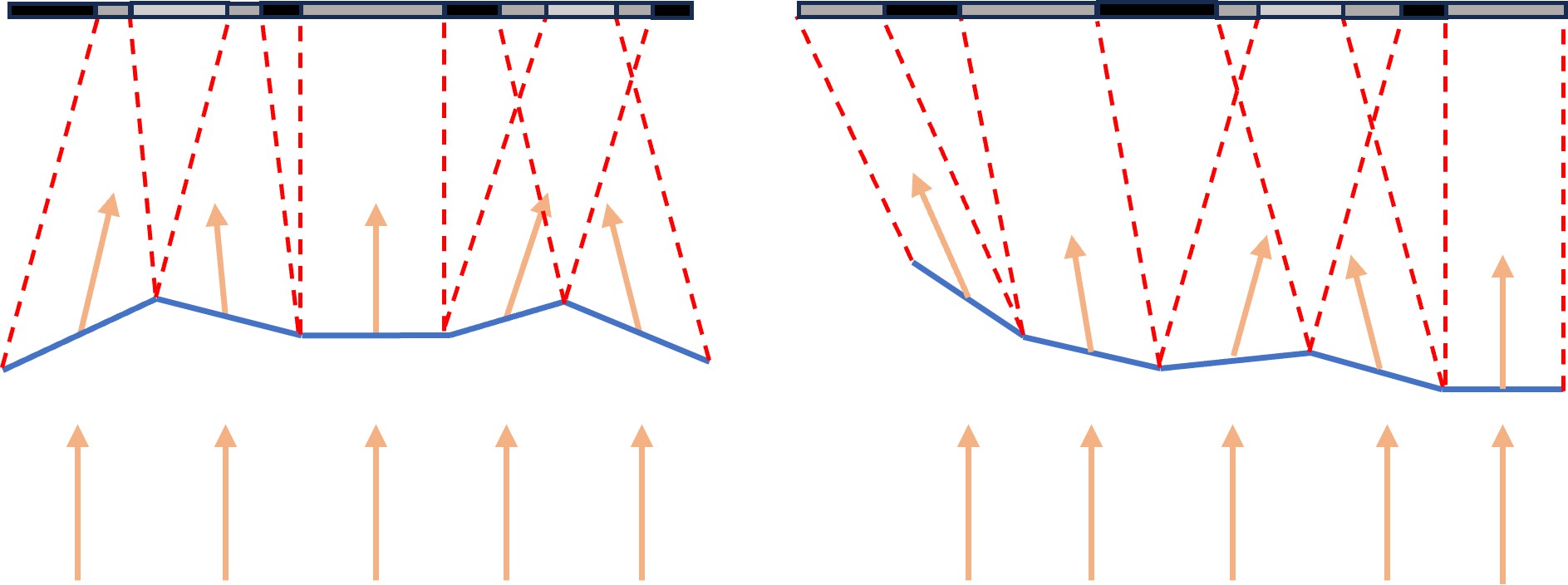}
  \caption{\label{refraction_render}
           The refraction process on the back surface of lenses with different shapes and the rendering process on the receiving plane. The area with more luminous flux distribution on the receiving plane will have higher brightness.}
\end{figure}

Assuming that the distribution of flux is uniform within any sufficiently small triangular region, we can allocate the source flux of the projected triangle $t_i''$ to each pixel $p^j$ on the plane based on the intersecting area, as shown in the following equation:
\begin{equation}
\Phi_{ij}^k = \Phi_i^k \frac{\mathcal{A}(t''_i \cap p^j)}{\mathcal{A}(t''_i)}.
\end{equation}
where $\Phi_{ij}^k$ represents the source flux contributed by the image triangle $t_i''$ to the pixel $p^j$, $\mathcal{A}(\cdot)$ denotes the area size of a region.
Therefore, the total flux $ \Phi^j $ of a pixel $ p^j $ is the sum of the flux assigned to this pixel by each projection triangle $t''_i$ on the back surface of the lens:
\begin{equation}
\label{pixel_flux}
\Phi^j = \sum_{k=1}^{N} \sum_{i=1}^{N_t} \Phi_{ij}^k,
\end{equation}
where $N_t$ is the number of triangles, $\Phi_{ij}^k$ is the flux allocated to pixel $p^j$ by the k-th point light source $l^k$ through the triangle $t''_i$.
Figure~\ref{refraction_render} provides an intuitive illustration.
We refer to $\Phi^j$ as the target flux on the receiving plane.
In order to reduce the cost of the rendering process, we only consider the projected triangles that intersect with each pixel in our calculations.
For each projection triangle $t''_i$, the scan-line algorithm can be used within its bounding box to determine which pixels it intersects with.
Therefore, Equation~\eqref{pixel_flux} can be further rewritten as:
\begin{equation}
\label{pixel_flux2}
\Phi^j = \sum_k \sum_{i \in I^j} \Phi_{ij}^k,
\end{equation}
where $I^j$ represents the index set of all projected triangles intersecting with $p^j$.

\paraheading{Gamma correction}
We have obtained the target flux of each pixel area on the receiving plane. Next, we will establish the relationship between it and the pixel grayscale value. Assuming the total brightness of the reference image is $G_{\tilde{\Phi}}$, then the grayscale value $\tilde{g}^j$ of pixel $p^j$ on the reference image satisfies:
\begin{equation}
\sum_{j=1}^{n_p} \gamma(\tilde{g}^j) = G_{\tilde{\Phi}},
\end{equation}
where $\gamma(\cdot)$ is the gamma correction, $n_p$ is the total number of pixels on the reference image.
Denote $\gamma_j = \gamma(g^j)$, where $g^j$ is the grayscale value at pixel $p^j$ in the rendered image. There is a proportional relationship between $\gamma_j$ and the target flux $\Phi^j$. Thus, we can derive that:
\begin{equation}
\label{flux2pixel}
g^j = \gamma^{-1} \bigg(G_{\tilde{\Phi}} \cdot \frac{\Phi^j}{\sum_{v=1}^{n_p} \Phi^v} \bigg)
\end{equation}
At this point, we have calculated the grayscale value of each pixel on the rendered image.

This rendering model accurately allocates luminous flux in a physically based way, eliminating the need for sampling light rays, thus significantly reducing computational costs and ensuring the differentiability of the entire process. In the next subsection, we will employ a gradient-based solver to optimize the variables of our light source model. 

\subsection{Optimization design for light source model}
After establishing the above differentiable rendering method, we can optimize our light source model based on its properties.
For a surface light source of a certain size, we give an appropriate numerical parameter $N$ and choose a relatively uniform initial distribution $\{(x_k, y_k) \mid k=1,2,\ldots,N\}$. 
The initial intensity parameters $q_k$ of each ideal point light source are equal.
By optimizing their positions and light intensity, we ensure that the model, after passing through a lens with a known shape, forms an image that closely matches the reference image on the receiving plane (see Figure~\ref{pipeline1}).
Here, the reference image refers to the caustic image formed after the surface light source passes through the input lens.
Next, we will introduce our philosophy and formulas in optimization design.

\paraheading{Reference data}
In order to enable the light source model to improve its expression ability of surface light source characteristics during the optimization process, we need to utilize two types of data: the input lens and its reference image.

To generate intricate lens shapes that can serve as effective references in the optimization process, we select target images with specific details and derive the input lenses through an optimization process that jointly considers flux consistency and normal integrability.\label{Reference_data}

After obtaining these input lenses, we generate their caustic images under a real surface light source using LuxCoreRender\footnote{\url{https://luxcorerender.org/}}, a physically-based renderer, referred to as reference images.
These reference images are the ground truth in the optimization process of our light source model. 

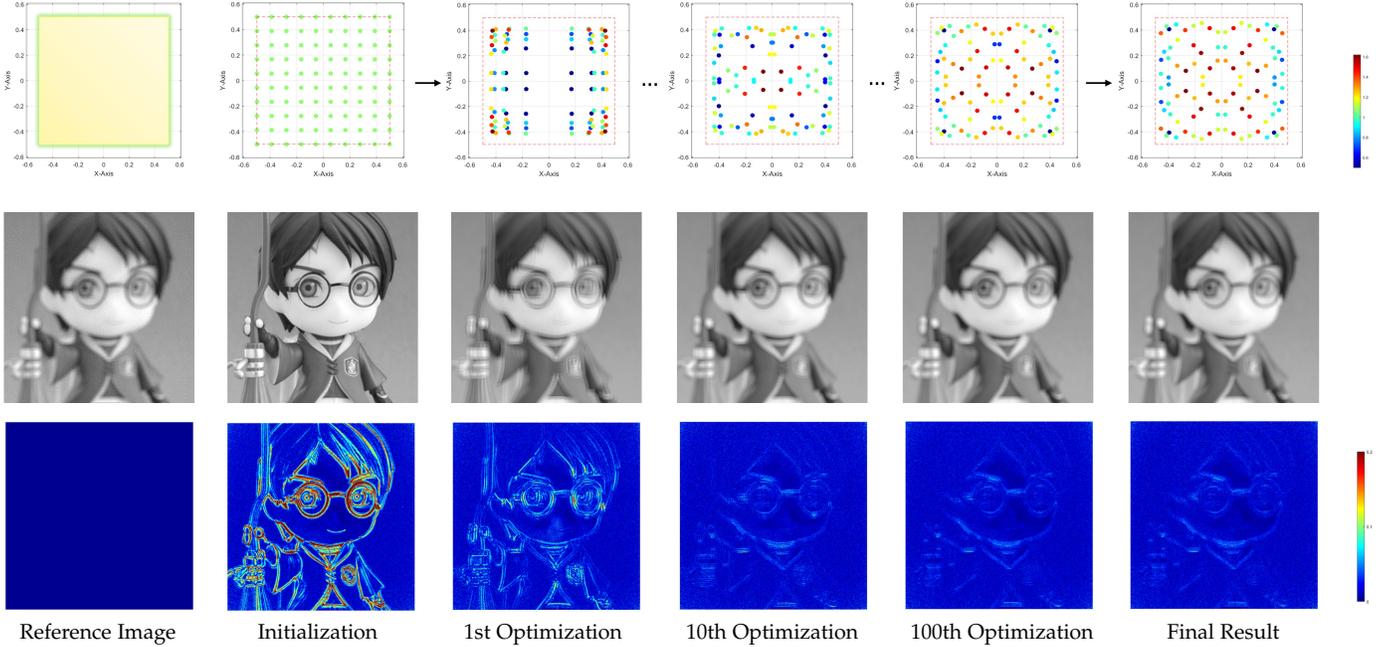
\begin{figure*}[tbp]
  \centering
  \mbox{} \hfill
  \hfill
  \includegraphics[width=1\linewidth]{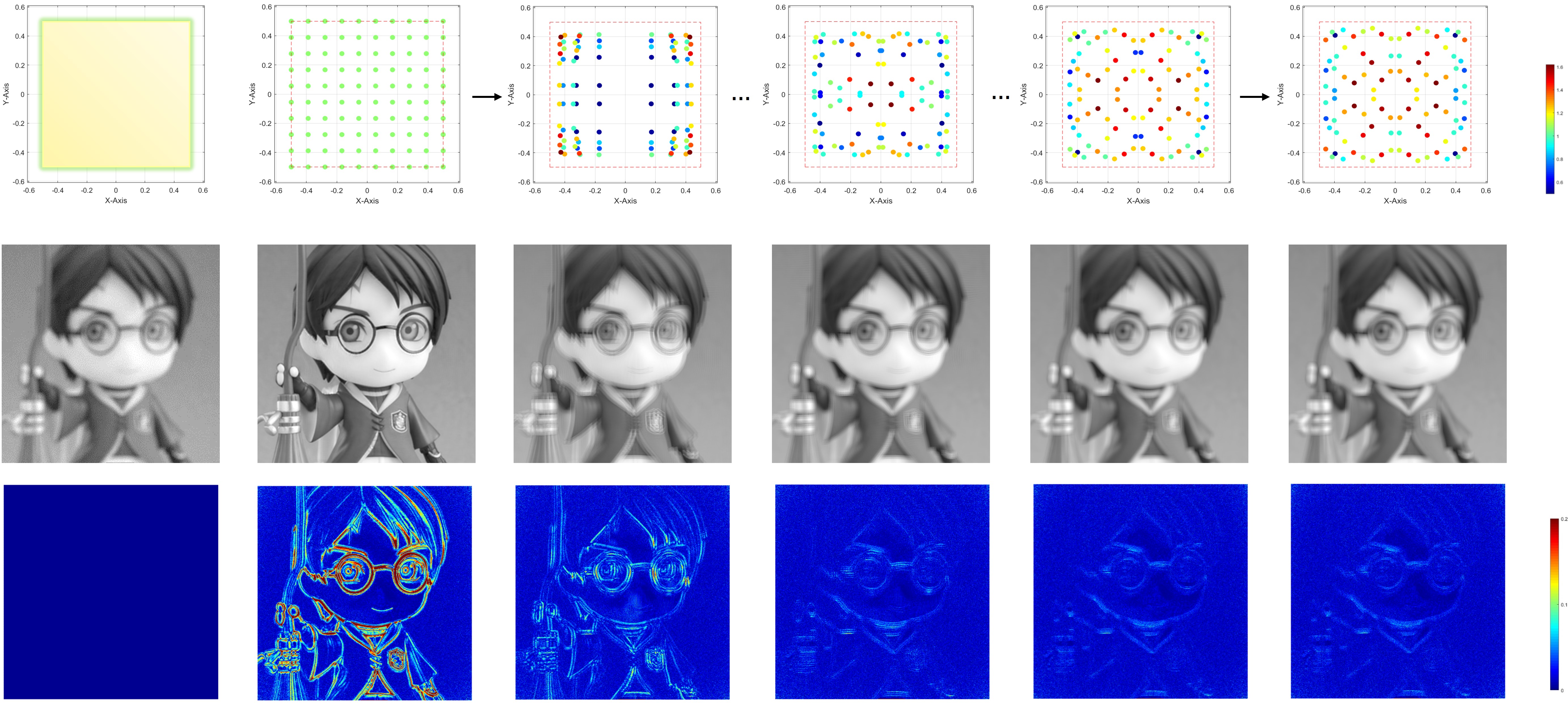}\\
  \vspace{-1em}
  \begin{picture}(0,0)
    \put(-251,-8){\footnotesize Reference Image}
    \put(-161,-8){\footnotesize Initialization}
    \put(-83,-8){\footnotesize 1st Optimization}
    \put(1,-8){\footnotesize 10th Optimization}
    \put(86,-8){\footnotesize 100th Optimization}
    \put(183,-8){\footnotesize Final Result}
  \end{picture}
  \vspace{0.6em}
  \caption{\label{ls_model}%
           The process of model initialization and optimization iterations. The first line is a visualized image of model parameters, $ x $ and $ y $ coordinates are the model's position parameters, while the color depth of the points indicates the light intensity parameters $ q $. Darker colors represent higher light intensities at those positions. The second line is the rendered result based on the current model. The third line is the color coding of the average pixel error between the rendered result and the observed image. The labels at the bottom display the current iteration count of the model.}
\end{figure*}

\paraheading{Optimization settings}
We take a rectangular light source with a side length of $B$ as an example. 
Assume that the plane where the surface light source, the front surface of the lens, and the receiving plane are located is parallel (all perpendicular to the $z$-axis of the spatial coordinate system), and the $x$ and $y$-coordinates of their centers are all zero.
The front surfaces of these lenses are rectangles with side lengths of $w \times h$. The back surfaces are composed of triangle meshes, and their boundaries are rectangular and identical to the front surfaces.
Denote the number of vertices of the triangle mesh on the boundary of the back surface as $w_v$ and $h_v$, respectively. 
The imaging area $U_t$ on the receiving plane is set to $[-W/2, W/2] \times [-H/2, H/2]$, and the resolution is $ W_p \times H_p$.
And we let the $z$-coordinate of the plane where the surface light source is located be zero, the $z$-coordinate of the planes where the front surface of the lens and the receiving plane are located be $z_{f}$, $z_{r}$. 

We use the following initialization method for the light source model to be optimized: for the position parameters $(x_k, y_k)$, we choose a relatively uniform initial distribution within the rectangular area, and at the same time make their initial intensity parameters all equal.
Since the rectangular light source has symmetry, we can further limit the position parameters of the model to be optimized in the first quadrant of the $x-y$ plane, and then obtain the information of the other three quadrants through symmetry.


\paraheading{Flux term}
To quantify the difference between the rendered image and the reference image, we introduce a term representing the mean squared error of the pixel flux values between the two images:
\begin{equation}
E_{\text{flux}} = \sum_{j=1}^{N_p} (\Phi^j - \tilde{\Phi}^j)^2,
\label{eq:fluxloss}
\end{equation}
where $\Phi^j$ and $\tilde{\Phi}^j$ are the target flux of the rendered image and the reference image at pixel $p^j$. 
Among them, $\tilde{\Phi}^j$ can be calculated by the following equation:
\begin{equation}
\tilde{\Phi}^j = \gamma(\tilde{g}^j)
\end{equation}
By penalizing $E_{\text{flux}}$, we can ensure that the rendered result becomes increasingly close to the reference image.

\paraheading{Optimization problem}
Next, we will present the optimization problem defined for the light source model. 
Before presenting the specific optimization problem, we first define two boundary penalty terms. 
These two boundary terms are respectively for the position parameters and intensity parameters of the light source model.
Since the position of each primitive $l^k$ in the light source model needs to be inside the actual surface light source, we define the position boundary term: \begin{equation}
E_{\text{bp}} = \sum_{k=1}^{N} \bigg( \sigma(|x_k| - \frac{B}{2})(|x_k| - \frac{B}{2})^2 + \sigma(|y_k| - \frac{B}{2})(|y_k| - \frac{B}{2})^2 \bigg),
\end{equation}
The definition of $\sigma(\cdot)$ here is as follows:
\begin{equation}
\sigma(x) = 
\left\{
\begin{aligned}
&\quad 1, \quad |x| > 0,
\\
&\quad 0, \quad |x| \leq 0.
\end{aligned}
\right.
\end{equation}
At the same time, since the intensity of $l^k$ is not less than zero, we define the light intensity boundary term: 
\begin{equation}
E_{\text{bi}} = \sum_{k=1}^{N} \big( \sigma(-q_k) \cdot q_k^2 \big).
\end{equation}
The introduction of these two hard constraints ensures that, after each optimization iteration, our light source model remains physically consistent (see Figure~\ref{ls_model}).
Thus, our optimization problem can be expressed as: 
\begin{equation}
\label{optimization problem1}
\min \quad E_{\text{total}} = \lambda_1 \sum_{m=1}^M E_{\text{flux}}^{(m)} + \lambda_2 E_{\text{bp}} + \lambda_3 E_{\text{bi}},
\end{equation}
where $\{\lambda_i\}$ are user-specified weights, $E_{\text{flux}}^{(m)}$ is the flux penalty term for the $m$-th set of reference data, and $M$ is the total number of input lenses.

\begin{figure*}[tbp]
  \centering
  \mbox{} \hfill
  \hfill
  \includegraphics[width=0.79\linewidth]{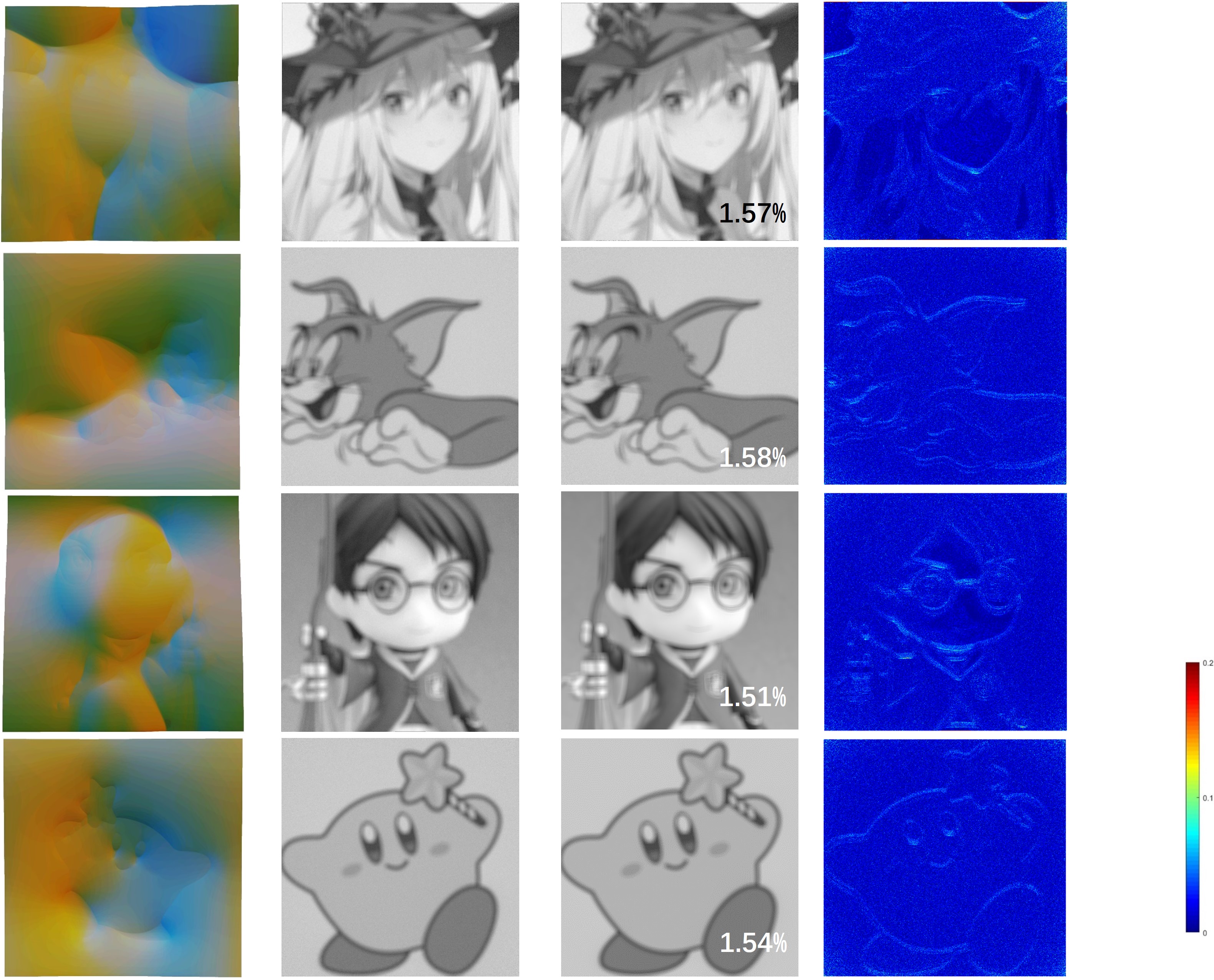}
  \hfill \mbox{}
   \begin{picture}(0,0)
    \put(-426,-12){\small Input Lens}
    \put(-346,-12){\small Reference Image}
    \put(-213,-12){\small Our Model(N = 64)}
  \end{picture}
  \vspace{1em}
  \caption{\label{ls_model_result}%
           The image rendered based on our light source model. The rendering results obtained by our method are very close to the observed images under the real surface light source.
           The number in the lower right corner of each image represents the MAE with the reference image, and the color-coding shows pixel-wise absolute error (see Equation~\eqref{absolute error}) compared to the reference image.}
\end{figure*}
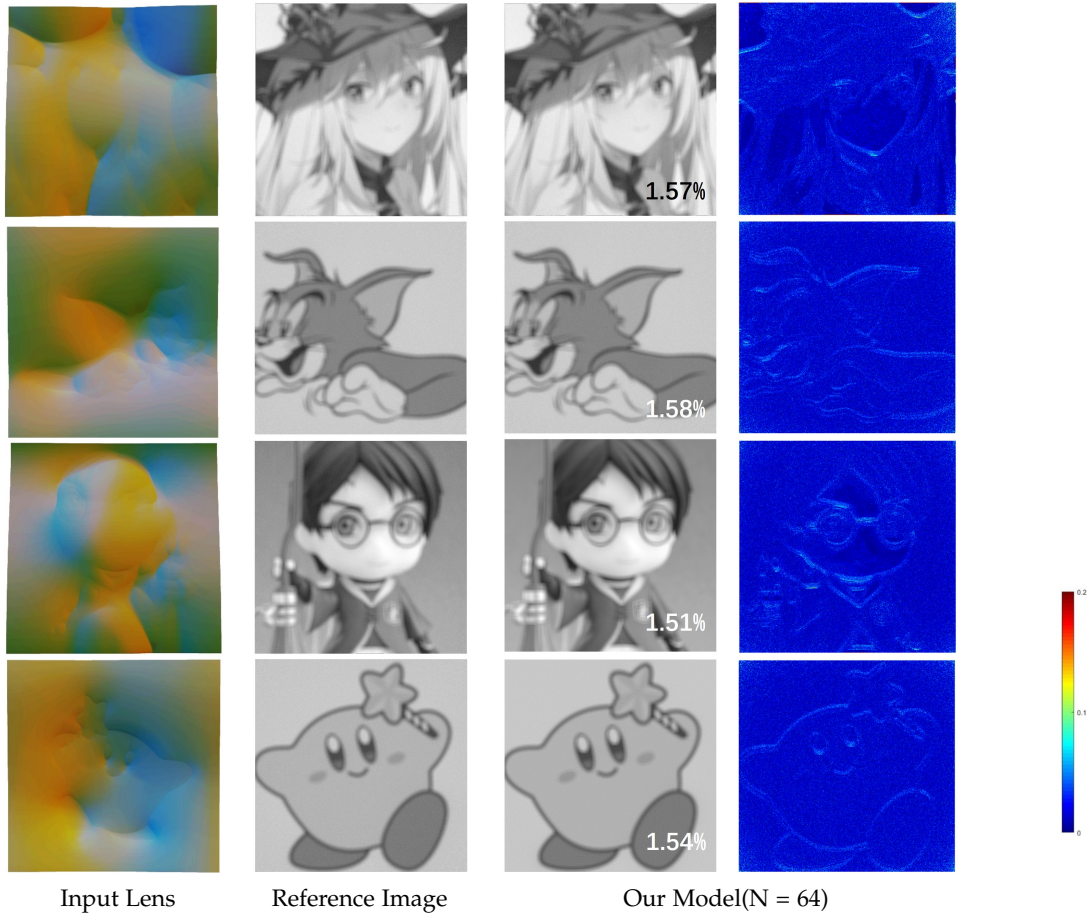

\begin{figure}[htb]
  \centering
  \includegraphics[width=1.0\linewidth]{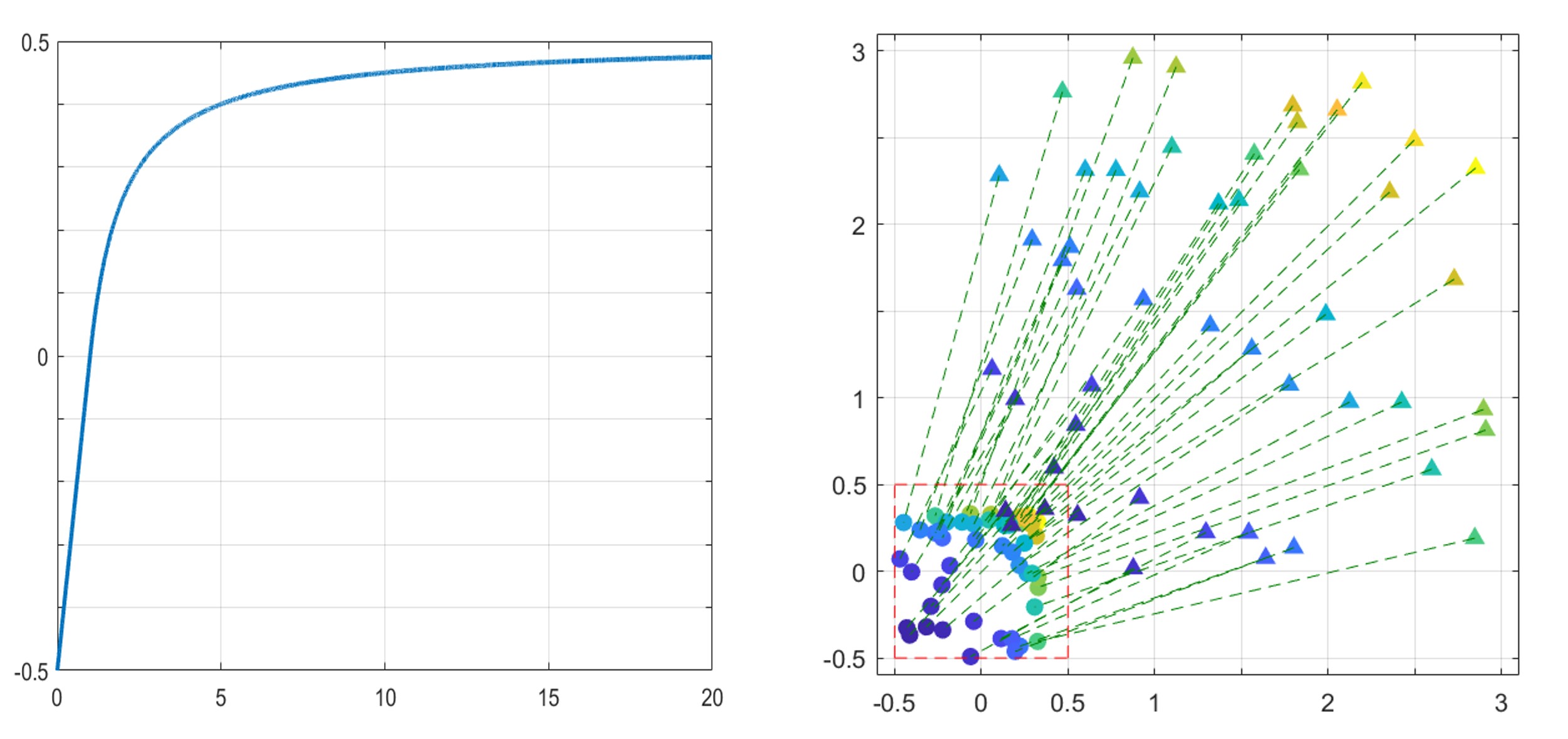}
  \begin{picture}(0,0)
    \put(-40,100){\small $\mathcal{T}(x)$}
  \end{picture}
  \vspace{-2em}
  \caption{\label{contraction_func}
           Left: contraction function curve. Right: mapping diagram of position parameters $(x_k, y_k)$ (assuming $B=1$ in the figures). The rectangular area in the lower left corner of the right figure is the physical boundary of our light source model.}
\end{figure}
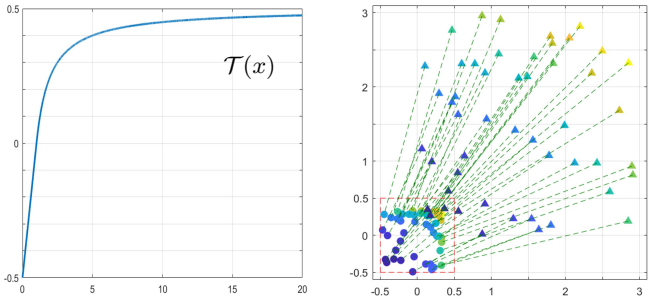

\paraheading{Contraction function}
In the above optimization problem~\eqref{optimization problem1}, the penalty for boundary constraints does not necessarily ensure that these parameters are strictly controlled within the boundaries. 
Furthermore, the selection of the coefficient of each penalty term has a significant impact on the optimization process and may lead to early termination.
Inspired by~\cite{barron2022mipnerf360}, we design a monotonic and continuously differentiable contraction function for the position parameters:
\begin{equation}
\mathcal{T}(x) = 
\left\{
\begin{aligned}
&\quad \frac{B}{2}(|x| - 1), \quad |x| \leq 1,
\\
&\quad \frac{B}{2}(1 - \frac{1}{|x|}), \quad |x| > 1.
\end{aligned}
\right.
\end{equation}
We regard the currently to-be-optimized parameters $(x_k, y_k)$ as intermediate variables and construct new variables $(\hat{x}_k, \hat{y}_k)$ that are directly optimized (see Figure~\ref{contraction_func}). 
The relationship between them satisfies: 
\begin{equation}
x_k = \mathcal{T}(|\hat{x}_k|), \quad y_k = \mathcal{T}(|\hat{y}_k|), \quad (\hat{x}_k, \hat{y}_k \in \mathbb{R}).
\end{equation}

Similarly, we also construct new optimization variables $\hat{q}_k \in \mathbb{R}$ for the light intensity parameters, which satisfy: ${q}_k = |\hat{q}_k|$.
Thus, the optimization problem~(\ref{optimization problem1}) is transformed into the following unconstrained optimization problem: 
\begin{equation}
\label{optimization problem2}
\min \quad \sum_{m=1}^{M} E_{\text{flux}}^{(m)}(\hat{X},\hat{Y},\hat{Q}) = \sum_m \sum_j(\Phi^j_m - \tilde{\Phi}^j_m)^2,
\end{equation}
where $\hat{X} = \{\hat{x}_k\}$, $\hat{Y} = \{\hat{y}_k\}$ and $\hat{Q} = \{\hat{q}_k\}$ are directly optimized parameters of the model, $\Phi^j_m$ and $\tilde{\Phi}^j_m$ respectively represent the target fluxes of the $m$-th group of rendered images and reference images.
It is worth emphasizing that since the contraction function has good differentiability, the entire optimization problem is fully differentiable throughout the process.
Through the above optimization process, the expression ability of the light source model for the illumination characteristics of the surface light source can be continuously improved.
It should be noted that, in the actual optimization process, we first perform a reverse contraction transformation $\mathcal{T}^{-1}(\cdot)$ on the initial values of $(x_k, y_k, q_k)$ to obtain the initial values of $(\hat{x}_k, \hat{y}_k, \hat{y}_k)$. 
After optimizing to convergence, the results are then mapped back to the actual region using the mapping $\mathcal{T}(\cdot)$. 

To solve unconstrained optimization problems~(\ref{optimization problem2}), we use the L-BFGS algorithm~\cite{Liu1989OnTL} to compute the numerical solution and leverage GPU parallel computation for the target value and its gradient (discussed in the next section). 
Due to the scale and complexity of the aforementioned optimization problem, we use an automatic differentiation data structure to precisely evaluate the gradient after each iteration.

\begin{figure*}[tbp]
  \centering
  \mbox{} \hfill
  \hfill
  \includegraphics[width=1\linewidth]{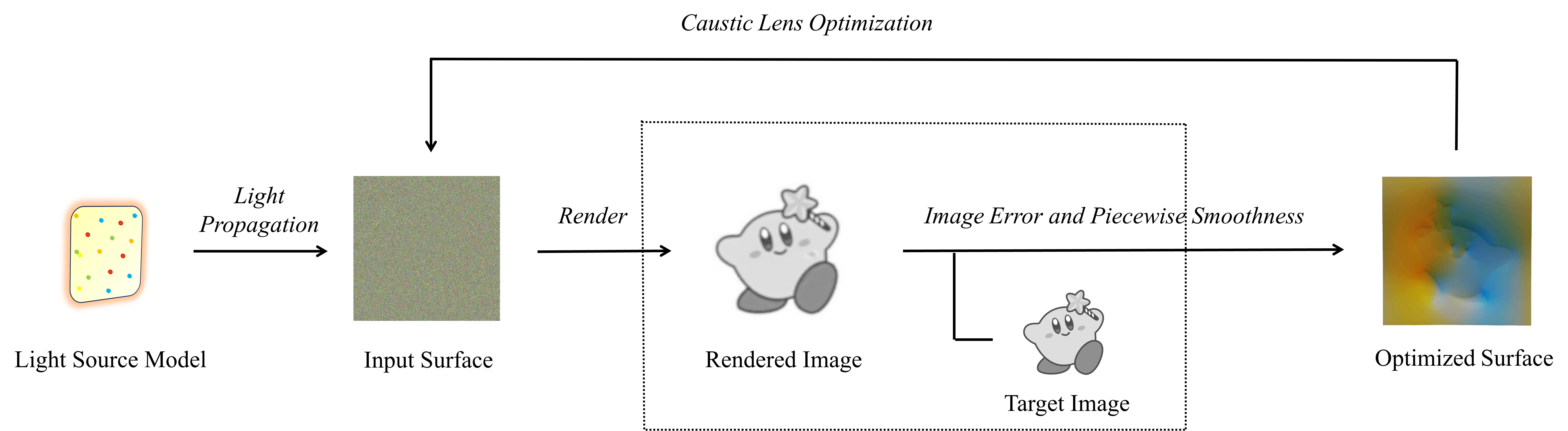}
  \hfill \mbox{}
  \vspace{-1.5em}
  \caption{\label{pipeline2}%
           Caustic lens optimization process: using the optimized light source model as input to guide the optimization of the caustic surface through image loss and smooth loss.}
\end{figure*}

\paraheading{Parallel design}
Due to the high computational costs for the rendering and optimization process, we have designed a parallel algorithm to improve the solution efficiency.
The scale of the optimization problem~(\ref{optimization problem2}) mainly depends on the size of the number parameter $N$ of the light source model and the number of triangle meshes on the back surface of the input lens.
It is easy to find that in the actual optimization process, the latter is often much larger than the former. 
Therefore, we take each triangle mesh on the back surface as a thread for parallel design.

Another important factor affecting the efficiency of the algorithm is the calculation of the gradient. 
The gradient of the optimization problem~(\ref{optimization problem2}) can be obtained by the following formula: 
\begin{equation}
\frac{\partial E}{\partial v} = \sum_m \sum_j \frac{\partial}{\partial v} (\Phi^j_m(\hat{X},\hat{Y},\hat{Q}) - \tilde{\Phi}^j_m)^2,
\end{equation}
where $v \in \{x_k\} \cup \{y_k\} \cup \{q_k\}$.
To solve these gradient values, we use an automatic differentiation data structure in the algorithm. 
Therefore, each step involves the calculation of $3N$ partial derivatives (each intermediate variable is $3N + 1$ dimensional, including variable values and partial derivative values), which incurs huge time and space overhead.
To solve this problem, we make full use of the characteristics of this light source model and propose an algorithm that can significantly reduce the computational cost.
We re-express the optimization problem~(\ref{optimization problem2}) as: 
\begin{equation}
\label{optimization problem3}
\min \quad E(\hat{X},\hat{Y},\hat{Q}) = \sum_m \sum_j(\sum_i \sum_k \Phi^{k,m}_{ij} - \tilde{\Phi}^j_m)^2,
\end{equation}
where $\Phi^{k,m}_{ij}$ represents the flux allocated by $l_k$ to pixel $p_j$ of the $m$-th rendered image through $t'_i$.
With the help of the chain rule, we can infer that: 
\begin{equation}
\label{partial derivative}
\begin{aligned}
\frac{\partial E}{\partial v_s} =  &~ 2 \sum_m \sum_j (\Phi^j_m - \tilde{\Phi}^j_m) \cdot \frac{\partial}{\partial v_s} \bigg( \sum_i \sum_k \Phi^{k,m}_{ij} (\hat{X},\hat{Y},\hat{Q}) \bigg)
\\
=  &~ 2 \sum_m \sum_j (\Phi^j_m - \tilde{\Phi}^j_m) \cdot \sum_i \frac{\partial}{\partial v_s} \Phi^{s,m}_{ij} (x_s, y_s, q_s),
\end{aligned}
\end{equation}
where $v_s \in \{x_s, y_s, q_s\}$.
This takes advantage of the independence between each primitive in the light source model.
Therefore, we only need to store three partial derivatives in the automatic differentiation structure (not related to the size of $N$).
Accordingly, we need to additionally add a parallel process of summing the partial derivatives $\Phi^{s,m}_{ij} (x_s, y_s, q_s)$. 
Through the above design, we greatly reduce the computational overhead of the overall optimization by only adding a small-scale parallel process.
In the next section, we will embed the optimized light source model into the lens optimization process.

\subsection{Caustic design based on optimized model}
After obtaining the optimized light source model, our current goal is to optimize the lens shape so that the caustic images formed under the combination of these point light sources are closely aligned with the given target images (see Figure~\ref{pipeline2}).
Given the high complexity and nonlinearity of this optimization, it would be challenging to optimize the lens shape directly from a planar surface.
We comprehensively consider flux consistency and normal integrability to obtain the initial shape of the lens.
Then, for optimization, we adapt the design of several loss terms from~\cite{sun2025end}.

\paraheading{Image term}
In order to make the rendering results close to the target image, we introduce a penalty term for the difference in their pixel values: 
\begin{equation}
\label{Image term}
E_{\text{img}} = \sum_{j=1}^{N_p} (g^j - \tilde{g}^j)^2,
\end{equation}
where $g^j$ and $\tilde{g}^j$ are the pixel values of the rendered image and the target image at pixel $p^j$.

The transition to a pixel-based error metric for this stage, in contrast to the flux-based metric used for light source optimization (Equation~\eqref{eq:fluxloss}), reflects the different objectives of the two optimization phases. While the light source optimization aims to match the physical flux distribution as accurately as possible, the goal of caustic design is to produce an image that is visually similar to the target. It is well-known that human perception of brightness does not scale linearly with physical flux values. The pixel grayscale values, which incorporate gamma correction (as described in Section~\ref{diff_rendering}), provide a much better approximation of perceived brightness. Therefore, optimizing on pixel differences allows our method to prioritize perceptually significant improvements in the final caustic image.

Meanwhile, in order to preserve the prominent features in the target image, we construct a penalty term for the difference between their image gradients:
\begin{equation}
E_{\text{grad}} = \| G_x - \widetilde{G}_x \|_F^2 + \| G_y - \widetilde{G}_y \|_F^2,
\end{equation}
where $G_x, G_y$ and $\widetilde{G}_x, \widetilde{G}_y$ represent the gradient matrices of the rendered image and target images, respectively. 

In addition, we also introduce a light energy loss term to penalize the situation where light rays are refracted outside the specified imaging area: 
\begin{equation}
E_{\text{out}} = \sum_{i=1}^{n_t} \sum_{k=1}^3 \| \widetilde{v}_i^k - P_I(\widetilde{v}_i^k) \|^2,
\end{equation}
where $\widetilde{v}_i^k \in \mathbb{R}^2$ is $k$-th vertex position of the projected triangle $t''_i$, and $P_I(\widetilde{v}_i^k) \in \mathbb{R}^2$ is the closest point to $\widetilde{v}_i^k$ from the imaging region.

\paraheading{Smooth term}
Compared to an optical surface design with a smooth shape, rough surfaces are very difficult to machine.
Moreover, polishing a rough surface is also a major challenge, especially on a surface represented as a high-resolution mesh.
To make the fabricated lens closer to the precision in the design stage, we hope to obtain freeform surfaces with better overall smoothness. 
Therefore, a smoothness term that penalizes the mean curvature is introduced:
\begin{equation}
E_{\text{smooth}} = \sum_{i=1}^{n_t} (H_i)^2 A_i,
\end{equation}
where $H_i$ is the discrete mean curvature for the face $t_i$, $A_i$ is the area of $t_i$. 
Specific details can be referred to~\cite{sun2025end}.

\paraheading{Optimization problem}
Combining all the above loss terms, we present the final surface optimization problem:
\begin{equation}
\label{optimization problem_caustic}
\min \quad \mu_1 E_{\text{img}} + \mu_2 E_{\text{grad}} + \mu_3 E_{\text{out}} + \mu_4 E_{\text{smooth}},
\end{equation}
where $\{\mu_i\}$ represents the weights of each loss term. 
To maintain consistency, we still adopt the same solver and the design of the automatic differentiation structure as those used in optimizing the light source model.

\section{Experiments}
In this section, we will first present some specific details of the experimental setup and the evaluation metrics. 
Subsequently, we will display the optimization results of the light source model to verify its effectiveness.
Based on the optimized model, we provide the shape optimization results of the freeform surface lens.
Finally, we present the real rendering results of the physical prototypes of the corresponding lenses.

\subsection{Implementation Details}

\frontparaheading{Light source model}
In the actual computation, taking into account both computational costs and model performance, we choose $M = 4$ as the number of groups of reference data. 
We set the side length of the rectangular light source $ B = 1 $ cm. 
The front surface of the lens is a rectangle with side lengths $ w = 10 $ cm and $ h = 10 $ cm. 
The back surface of the lens has $ w_v = 641 $ and $ h_v = 641 $ vertices on the boundary. 
The width and height of the reference image $ W = 9.9 $ cm, $ H = 9.9 $ cm. 
And the resolution of images is set to $ 640 \times 640 $. 
The $z$-coordinates of planes where the front surface of the lens and the receiving plane are located are $120$cm and $150$cm, respectively. The refractive index $\eta$ of the lens is $1.49$. 
For the number parameter of point light sources that constitute the overall light source model, we select $N = 36$, $64$, and $100$, respectively.

\paraheading{Caustic lens}
In caustic design, we let the light source be parallel to the front surface of the lens and the receiving plane. 
The distance between the plane of the surface light source and the front surface of the lens is $120$ cm.
The distance between the light source and the receiving plane is $240$ cm.
The resolution of the rendered image is $ 640 \times 640 $ with an actual size of $ 20.0 $ cm $\times 20.0 $ cm. 
The size of the light source model is set to $0.1$ cm. 
The size of the front surface of the lens is $10 $ cm $\times 10 $ cm. The triangle mesh representing the back surface of the lens consists of 1,638,400 triangles.
The refractive index $\eta$ of the lens is $1.49$. 

\paraheading{Optimization}
In the optimization process of the surface light source model, we set the termination condition as the norm of the current gradient of the objective function being less than $10^{-2}$. In the optimization process of caustics design, the termination condition is met if either the norm of the current gradient of the objective function is less than $10^{-4}$, or the total number of iterations reaches $300,000$.  All the rendering and optimization processes are implemented on the GPU. In the optimization stage of the light source model, we use NVIDIA GeForce RTX 3090 GPUs, and in the caustic design stage, we use NVIDIA H100 GPUs.

\subsection{Evaluation metrics}
In the process of light source modeling and caustic design, to verify the effectiveness of our method, we introduce two evaluation metrics: mean absolute error (\textit{MAE}) and peak signal-to-noise ratio (\textit{PSNR}). 

\textit{MAE} refers to the average of the absolute errors between two objects, which can directly reflect the degree of closeness between two sets of data.
In Table~\ref{table1} and Table~\ref{table2}, we evaluate the difference between two images by using (\textit{MAE}):
\begin{equation}
\textit{MAE} = \bigg( \sum_{j=1}^{n_p} e_j \bigg) / n_p,
\end{equation}
where $e_j$ is the pixel-wise absolute error defined as:
\begin{equation}
\label{absolute error}
e_j = |p^j - q^j| / C_{\max},
\end{equation}
$ p^j $ and $ q^j $ are the pixel values of the rendered image and the target image, $ C_{\max} $ is the maximum possible pixel value of images.

Additionally, we also use the peak signal-to-noise ratio to evaluate the similarity of the images.
\textit{PSNR} is a full-reference quality evaluation metric. 
It can reflect to some extent the differences that are not directly observable by the human eye, such as image blurring, noise, block effects, and other distortion phenomena:
\begin{equation}
\textit{PSNR} = 10 \cdot \log_{10} \bigg( \frac{C_{\max}^2}{\textit{MSE}} \bigg),
\end{equation}
where \textit{MSE} is given by the following formula:
\begin{equation}
\textit{MSE} = \frac{1}{H_p \times W_p} \cdot \sum_{j=1}^{N_p} (p^j - q^j)^2.
\end{equation}

\subsection{Surface light source model}
Recall that when light emitted by a surface light source passes through a lens, a blurred image is formed on the receiving plane. 
This reflects its unique illumination characteristics.
However, in many computational rendering scenarios, it is usually assumed that the light source is a point light source or parallel light, which is obviously inaccurate.
Therefore, we hope to model the real surface light source and optimize the model parameters based on the rendering model, so that the light source model can reflect the real characteristics of the physical light source.

\begin{table}[t]
\belowrulesep=0pt
\aboverulesep=0pt
\centering
\renewcommand{\arraystretch}{1.5}
\caption{The \textit{MAE} and \textit{PSNR} of the rendered images with the reference image for both the un-optimised and optimised models through four freeform lenses. 
\textit{MAE}$^{0}$ and \textit{PSNR}$^{0}$ are obtained from the model with initial parameters, whereas \textit{MAE}$^{*}$ and \textit{PSNR}$^{*}$ come from the model after optimisation converges.}

\begin{tabular}{
  >{\centering\arraybackslash}m{1.2cm} |
  >{\centering\arraybackslash}m{0.9cm}
  >{\centering\arraybackslash}m{0.9cm}
  >{\centering\arraybackslash}m{1.3cm}
  >{\centering\arraybackslash}m{0.9cm}
  >{\centering\arraybackslash}m{0.9cm}}
\toprule
Image & $N$ & MAE$^{0}$ ($\times 10^{-2}$) & MAE$^{*}$ ($\times 10^{-2}$)  $\downarrow$& PSNR$^{0}$ (dB) & PSNR$^{*}$ (dB) $\uparrow$\\
\midrule
\multirow{3}{*}{Skadii}
 & 36  & 4.320 & 2.240 & 23.61 & 30.03\\
\cmidrule(r){2-6}
 & 64  & 4.397 & 1.574 & 23.44 & 32.52\\
\cmidrule(r){2-6}
 & 100 & 4.441 & 1.532 & 23.36 & 32.69\\
\midrule
\multirow{3}{*}{Tom}
 & 36  & 3.792 & 2.160 & 23.34 & 30.60\\
\cmidrule(r){2-6}
 & 64  & 3.878 & 1.578 & 23.15 & 33.95\\
\cmidrule(r){2-6}
 & 100 & 3.870 & 1.565 & 23.07 & 34.03\\
\midrule
\multirow{3}{*}{Harry}
 & 36  & 3.759 & 2.117 & 24.04 & 30.61\\
\cmidrule(r){2-6}
 & 64  & 3.825 & 1.508 & 23.88 & 33.96\\
\cmidrule(r){2-6}
 & 100 & 3.827 & 1.457 & 23.81 & 34.27\\
\midrule
\multirow{3}{*}{Kirby}
 & 36  & 3.025 & 1.915 & 25.49 & 31.75\\
\cmidrule(r){2-6}
 & 64  & 3.101 & 1.539 & 25.34 & 34.16\\
\cmidrule(r){2-6}
 & 100 & 3.057 & 1.533 & 25.33 & 34.20\\
\bottomrule
\end{tabular}
\label{table1}
\end{table}

Table~\ref{table1} reflects the effectiveness of our model.
The \textit{MAE} value indicates that the rendered image generated by our proposed light source model after optimization is very close to the reference image, as shown in Figure~\ref{ls_model_result}.
In addition, we visualized the changes in the model's parameters during the optimization process (see Figure~\ref{ls_model}).
As the number of iterations increases, the rendered image gradually approaches the reference image, which verifies the effectiveness of the optimization algorithm.

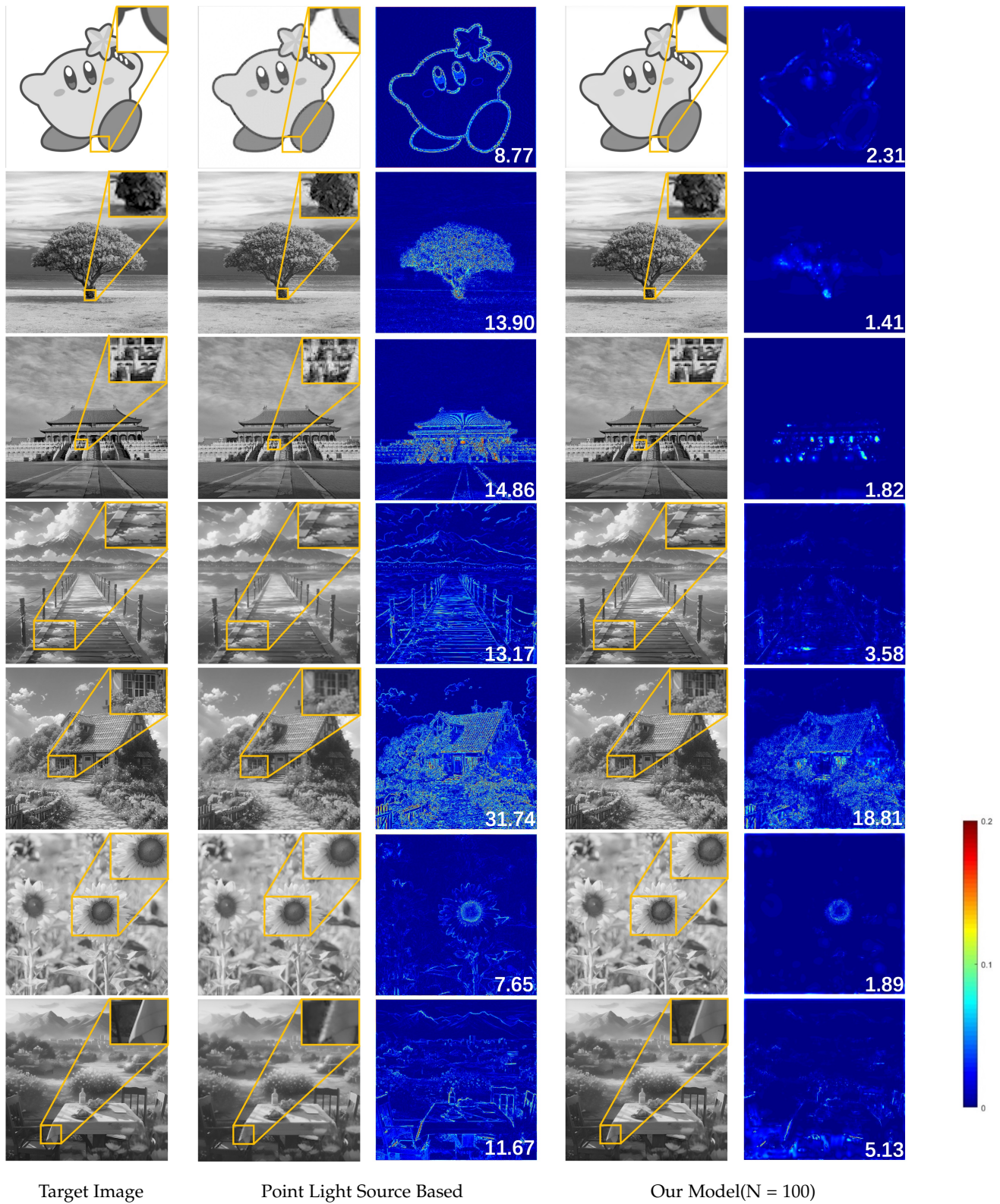
\begin{figure*}[tbp]
  \centering
  \mbox{} \hfill
  \hfill
  \includegraphics[width=0.95\linewidth]{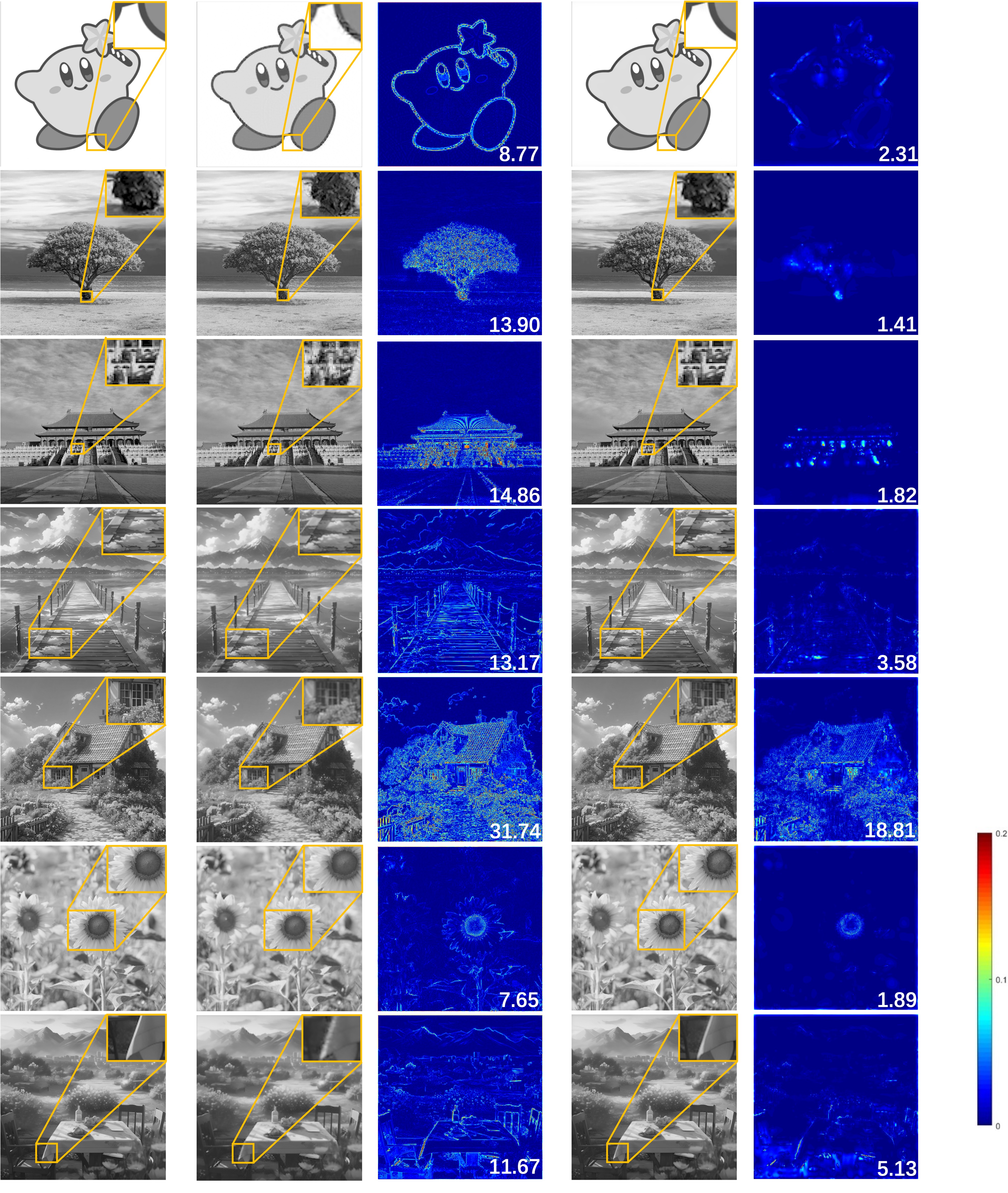}
  \hfill \mbox{}
  \begin{picture}(0,0)
    \put(-486,-12){\small  Target Image}
    \put(-375,-12){\small  Point Light Source Based}
    \put(-181,-12){\small  Our Model(N = 100)}
  \end{picture}
  \vspace{1.5em}
  \caption{\label{caustic_result}
           \textbf{The first column:} Some target images with rich details. \textbf{The second column:} Rendering results of the lens obtained by caustic design based on point light sources under the surface light source. \textbf{The third column:} Color-coding of the mean absolute error (\textit{MAE}) between the rendering images in the second column and the target images. \textbf{The fourth column:} Rendering results of the lens obtained by caustic design based on our model under the surface light sources. \textbf{The fifth column:} Color-coding of the \textit{MAE} between the rendering images in the fourth column and the target images. All error maps are processed using the same method. The number in the lower right corner of each image represents the \textit{MAE} ($\times10^{-3}$) with the target image. Some regions are enlarged for better inspection of details. }
\end{figure*}

\subsection{Caustic design}
Given the input light source, target image, and refractive index of the lens, caustic design is achieved by optimizing the shape of the lens to control the direction of light, resulting in an image on the receiving plane that is close to the target image. 
In the rendering process, the light source is usually assumed to be an ideal point light source, which is not physically accurate. 
Specifically, if the lens is optimized based on the illumination mode of a point light source, a blurred observed image will be formed under the actual light source.

We obtain the caustic lens corresponding to the target image based on the point light source and our light source models respectively.
Figure~\ref{caustic_result} shows their caustic effects under a surface light source. 
We can intuitively find that the caustic design based on the optimized light source model obtains a rendering result that is very close to the target image and is also relatively accurate in details.
At the same time, combined with the results in Table~\ref{table2}, we verify the correctness of the overall method.
Table~\ref{table_run} shows some specific situations during the optimization process of different lenses.

Our analysis ignores errors from reflection, absorption, and surface imperfections, an assumption justified by our use of acrylic (PMMA) for the physical prototypes. This material was chosen for its ideal optical and mechanical properties. In the visible spectrum, acrylic exhibits very low reflectance and near-zero absorption, which minimizes energy loss during light transport. Furthermore, its excellent wear resistance and machinability allow for a high-fidelity surface finish that reduces the impact of fabrication-related imperfections~\cite{PMMA2016review, PMMA2014Study}. Since these secondary physical effects are demonstrably minimal in our experimental setup, it is a valid and well-justified simplification for our model to focus on controlling light transport purely through the refractive properties of the lens geometry.

\begin{table}[t]
\belowrulesep=0pt
\aboverulesep=0pt
\centering
\renewcommand{\arraystretch}{1.5}
\caption{The \textit{MAE} and \textit{PSNR} of the rendered images with the target images under surface light source illumination for caustic design, which respectively used a point light source model and our rectangular surface light source model (N=100).}

\begin{tabular}{
    >{\centering\arraybackslash}m{1.4cm} |
    >{\centering\arraybackslash}m{2.0cm} 
    >{\centering\arraybackslash}m{1.6cm} 
    >{\centering\arraybackslash}m{1.6cm}}
\toprule
Image & Light source & MAE ($\times10^{-3}$) $\downarrow$ & PSNR (dB) $\uparrow$\\
\midrule
\multirow{2}{*}{Kirby} & Point-based   & 8.7695          & 31.4568\\
\cmidrule(r){2-4}
                       & Ours & \textbf{2.3148} & \textbf{44.2671}\\
\midrule
\multirow{2}{*}{Tree}  & Point-based   & 13.908          & 29.6808\\
\cmidrule(r){2-4}
                       & Ours & \textbf{1.4142} & \textbf{46.0483}\\
\midrule
\multirow{2}{*}{Palace}& Point-based   & 14.865          & 28.8148\\
\cmidrule(r){2-4}
                       & Ours & \textbf{1.8217} & \textbf{41.9505}\\
\midrule
\multirow{2}{*}{Bridge}& Point-based   & 13.174          & 32.3940\\
\cmidrule(r){2-4}
                       & Ours & \textbf{3.5797} & \textbf{41.5131}\\
\midrule
\multirow{2}{*}{House} & Point-based   & 31.740          & 25.7199\\
\cmidrule(r){2-4}
                       & Ours & \textbf{18.812} & \textbf{29.7512}\\
\midrule
\multirow{2}{*}{Flower}& Point-based   & 7.6488          & 35.9808\\
\cmidrule(r){2-4}
                       & Ours & \textbf{1.8860} & \textbf{44.0043}\\
\midrule
\multirow{2}{*}{Table} & Point-based   & 11.670          & 33.1791\\
\cmidrule(r){2-4}
                       & Ours & \textbf{5.1339} & \textbf{38.7847}\\
\bottomrule
\end{tabular}
\label{table2}
\end{table}

\begin{table}[ht]
\centering
\caption{The total time  for optimizing different lenses and the average time for a single iteration. The GPU memory size occupied by these lenses during the optimization process is approximately 3305 MB.}
\renewcommand{\arraystretch}{1.2}
\begin{tabular}{ccccc}
\toprule
\textbf{Lens} &
\makecell{\textbf{Total time}\\(hours)} &
\makecell{\textbf{Avg. time} \\(s/iter)} & 
\makecell{\textbf{MAE} \\($\times10^{-3}$)} & 
\makecell{\textbf{PSNR} \\(dB)}  \\
\midrule 
Kirby   & 1.21 & 2.12 & 2.31 & 31.46 \\
Tree    & 2.13 & 2.63 & 1.41 & 46.05 \\
Palace  & 2.32 & 2.85 & 1.82 & 41.95 \\
Bridge  & 2.14 & 2.77 & 3.58 & 41.51 \\
House   & 2.62 & 3.10 & 18.8 & 29.8 \\
Flower  & 3.72 & 2.72 & 1.89 & 44.00 \\
Table   & 4.10 & 3.01 & 5.13 & 38.78 \\
\bottomrule
\end{tabular}
\label{table_run}
\end{table}

\begin{figure*}[tbp]
  \centering
  \mbox{} \hfill
  \hfill
  \includegraphics[width=1.0\linewidth]{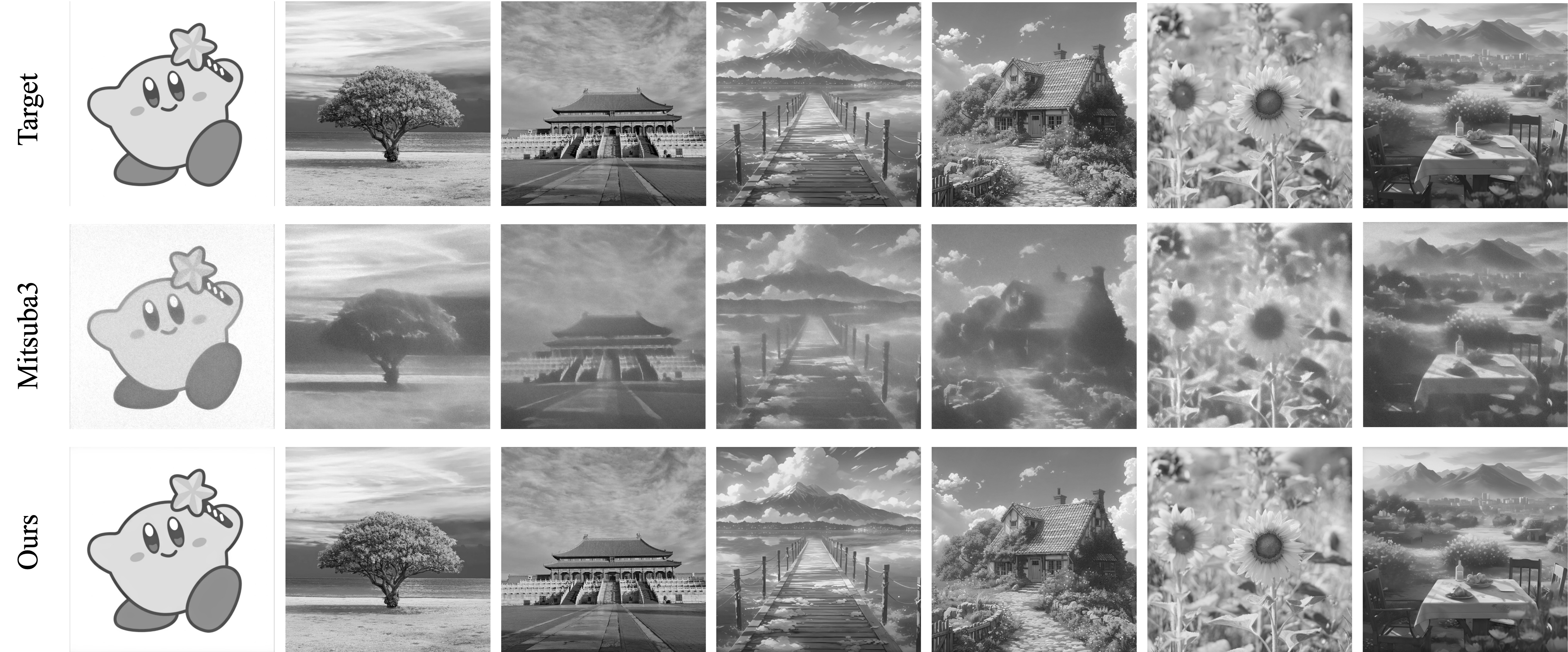}
  \hfill \mbox{}
  \vspace{-0.5em}
  \caption{\label{contrast_mitsuba3}%
           Comparison between our method and the rendered images produced by Mitsuba3. The rendering results of Mitsuba3 in caustic design exhibit a noticeable accuracy gap compared to our method from a visual perspective. In addition, its performance is suboptimal for caustic designs targeting images with high-contrast regions, such as areas containing pure white or pure black pixels.}
\end{figure*}

\subsection{Effectiveness of algorithm components}

\frontparaheading{Light transmission}
In the simulation of light transmission, we adopt a flux-based transmission mode. Compared with sampling light rays, our method has the following advantages.
First of all, each primitive in our transmission model has continuous illumination characteristics (flux has regionality), which is more accurate in a physical sense.
At the same time, compared to a large number of sampled light rays, our method obviously has a lower computational cost in the rendering process.
Most importantly, the sampling strategy needs to know the actual intensity distribution on the surface of the light source to accurately characterize its illumination characteristics, which is not easy in many scenarios.
However, our method does not need to determine any physical properties of the input light source in advance, so it has simplicity and universality.

We conducted comparative experiments with Mitsuba3~\cite{Mitsuba3}. 
Mitsuba3 uses a path tracing algorithm to simulate the light transport process, which is a variant of ray tracing.
It employs a Monte Carlo estimator to compute integral values by randomly emitting multiple rays to evaluate light intensity, thereby simulating the propagation of light in space.
This method requires sampling light rays, which has certain limitations in computational efficiency. Moreover, Mitsuba3 is not specifically designed for caustic design, and therefore its performance in terms of contrast and accuracy is inferior to our method (see Figure~\ref{contrast_mitsuba3}). 
It also completely neglects the consideration of actual fabrication in the overall design process.

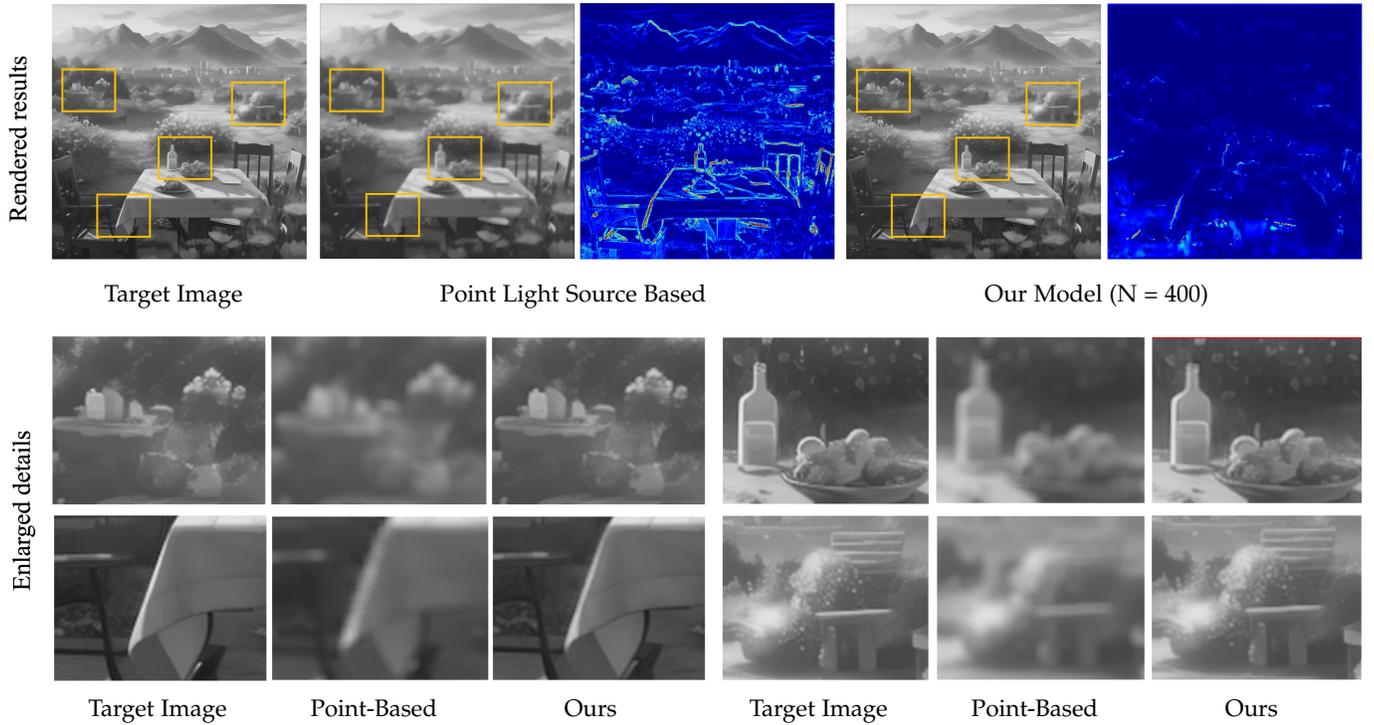
\begin{figure*}[tbp]
  \centering
  \mbox{} \hfill
  \hfill
  \includegraphics[width=1.0\linewidth]{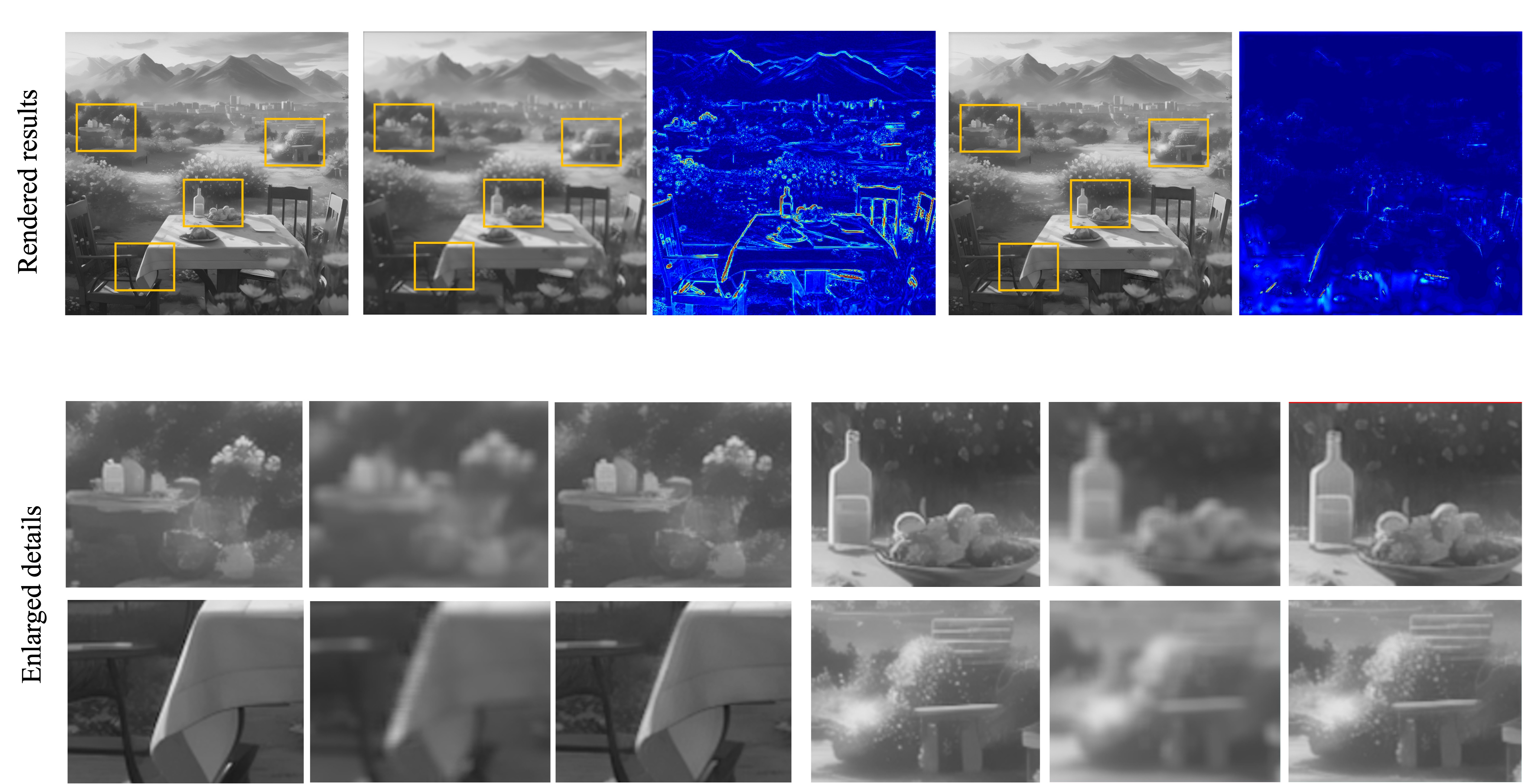}
  \hfill \mbox{}
  \vspace{-0.5em}
    \begin{picture}(0,0)
    \put(-217,155){\small  Target Image}
    \put(-90,155){\small  Point Light Source Based}
    \put(116,155){\small  Our Model (N = 400)}
    \put(-223,-2){\small Target Image}
    \put(-139,-2){\small Point-Based}
    \put(-43,-2){\small Ours}
    \put(27,-2){\small Target Image}
    \put(111,-2){\small Point-Based}
    \put(207,-2){\small Ours}
  \end{picture}
  \vspace{0.5em}
  \caption{\label{result_lslarge}%
        Comparison between the point light source based method and our method (N=400) under a surface light source with a side length of 0.2 cm. The first line shows the rendering results of the two methods, while the second and third lines provide a comparison of the details between these two methods. When the actual size of the surface light source is larger, the advantages of our method become more apparent.
    }
\end{figure*}

\begin{figure}[t]
  \centering
  \includegraphics[width=0.99\linewidth]{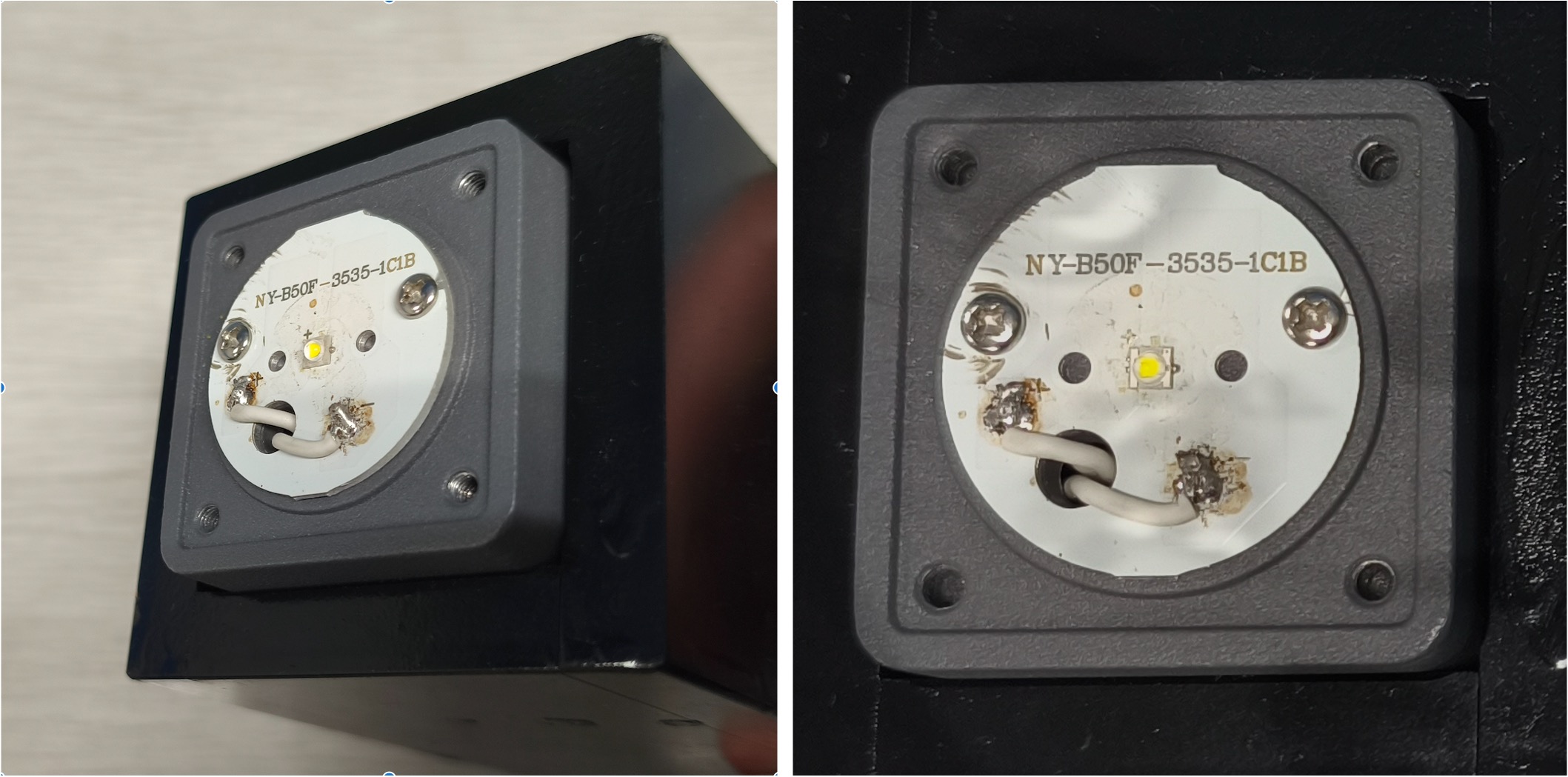}
  \caption{\label{physical_ls}
           The surface light source used in the physical experiment. The side length of it is roughly 1 mm. Its output power is 50W, and it is driven by a direct current of 12V and 4.2A.}
\end{figure}

\begin{figure}[t]
  \centering
  \includegraphics[width=1.0\linewidth]{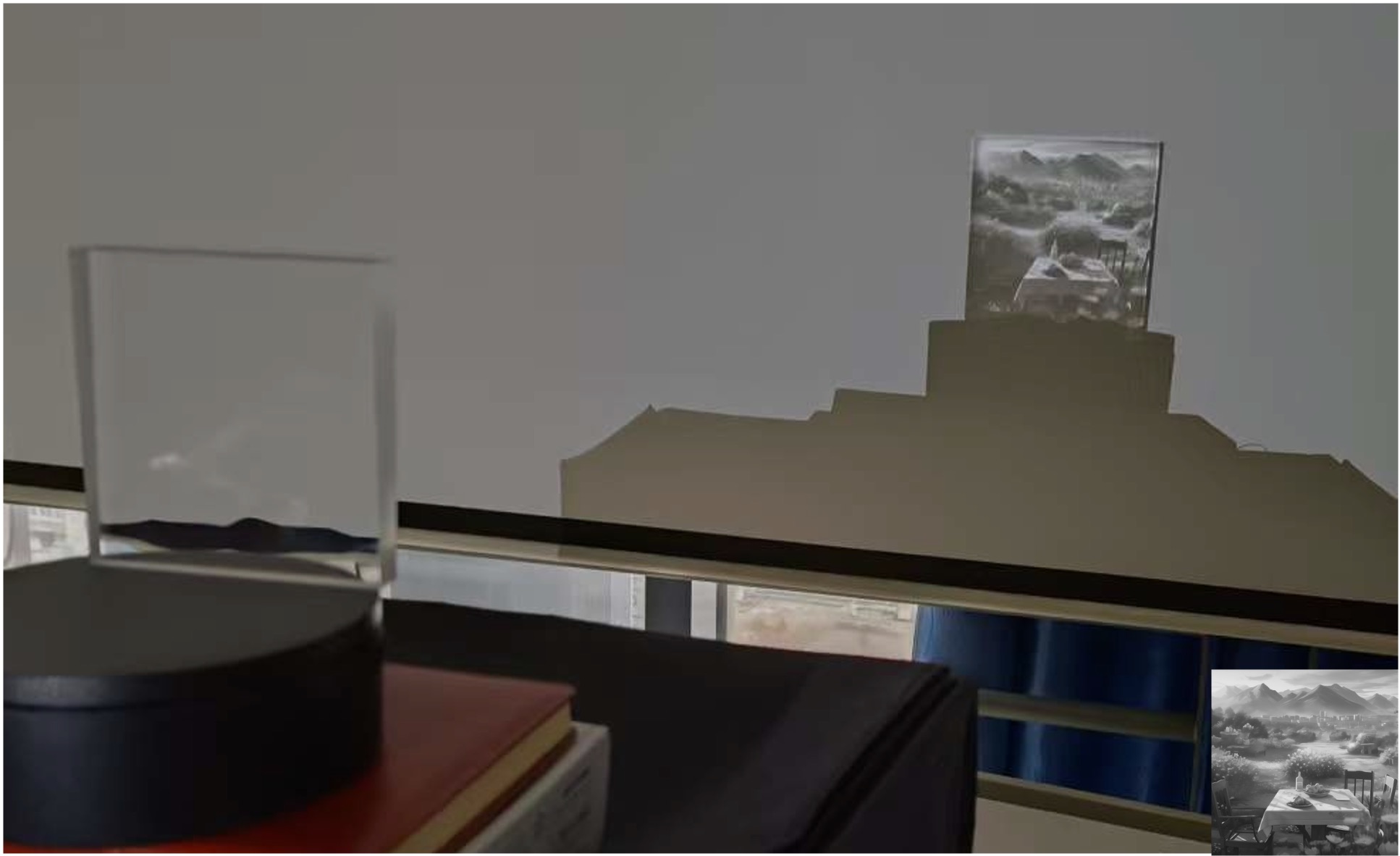}
  \caption{\label{physical_caustic}
           The physical prototype of the lens shape designed based on our method. Although some machining errors are introduced during the manufacturing process of the physical object, the real image generated by the prototype is still quite close to the image produced by our rendering model. Lower right corner: The target image corresponding to the lens.}
\end{figure}


\paraheading{Light source model}
In caustic scenes, it is not accurate enough to use a point light source to represent an actual LED surface light source.
Especially when the actual light source is relatively large in size or far away from an object, it will significantly affect the accuracy of lens shape calculation (see Figure~\ref{result_lslarge}).
Our light source model solves this problem.
Table~\ref{table2} gives a comparison between our method and the point light source based strategy.
Obviously, performing caustic design based on our optimized light source model can obtain rendering images that are closer to the target compared with a point light source.
In the seven examples provided, our approach achieved an average reduction of 71\% in the \textit{MAE} metric and an average improvement of 9.87 dB in the \textit{PSNR} metric.
It can also be found from Figure~\ref{caustic_result} that our method can significantly improve the quality of the rendered image, and at the same time, it also has better effects in some details.

Furthermore, our modeling method for light sources has good scalability and is directly applicable to curved surface light sources in three-dimensional space.
For LED light sources with different sizes but the same light intensity distribution, the corresponding model parameters can be obtained by means of scaling transformation, eliminating the need for repeated calculations.
In addition, our light source model only needs to be pre-optimized once. 
Once the model parameters are optimized, the pre-packaged model can be directly applied to rendering calculations in different scenarios.

A key methodological choice in our work is the discrete point source approximation. We considered an alternative, fully continuous formulation, but concluded that it is theoretically complex and computationally prohibitive for this application.
A continuous approach would require computing the flux from the entire surface light source onto each triangle of the lens's rear face. This necessitates evaluating a 2D integral over the light source area for every triangle on the lens, which presents two fundamental challenges.
Firstly, the integrand itself is non-trivial, as it must implicitly account for the first refraction at the lens's front surface. Deriving a closed-form, differentiable integral under these conditions is exceptionally difficult.
Secondly, for a high-resolution lens mesh like ours (with over 1.6 million triangles), this method would require performing millions of 2D integrations at each optimization step, leading to an intractable computational cost.
In contrast, our discrete model provides a pragmatic and effective solution. By approximating the source with a finite set of points, we avoid the complexity and cost of continuous integration. As demonstrated by our results (e.g., Figure~\ref{ls_model_result}), this approach still achieves the accuracy required to faithfully model the surface light source and produce high-quality caustic images.


\subsection{Physical verification}
We have presented the physical prototype of the lens designed using our method in Figure~\ref{physical_caustic}.
The material of the physical lens shown in this figure is acrylic (IOR: 1.49). 
The machining process is mainly carried out through a JDGR100 milling machine in a three-axis mode. 
Firstly, it undergoes rough milling and is then fabricated using a 0.4-mm ball-end mill. 
Finally, the lens needs to be manually polished with polishing wax and a cloth wheel.

During the experiment, we all used a rectangular surface light source with a side length of approximately 1 millimeter as the input light source, as shown in Figure~\ref{physical_ls}.

\begin{figure}[t]
  \centering
  \includegraphics[width=1.0\linewidth]{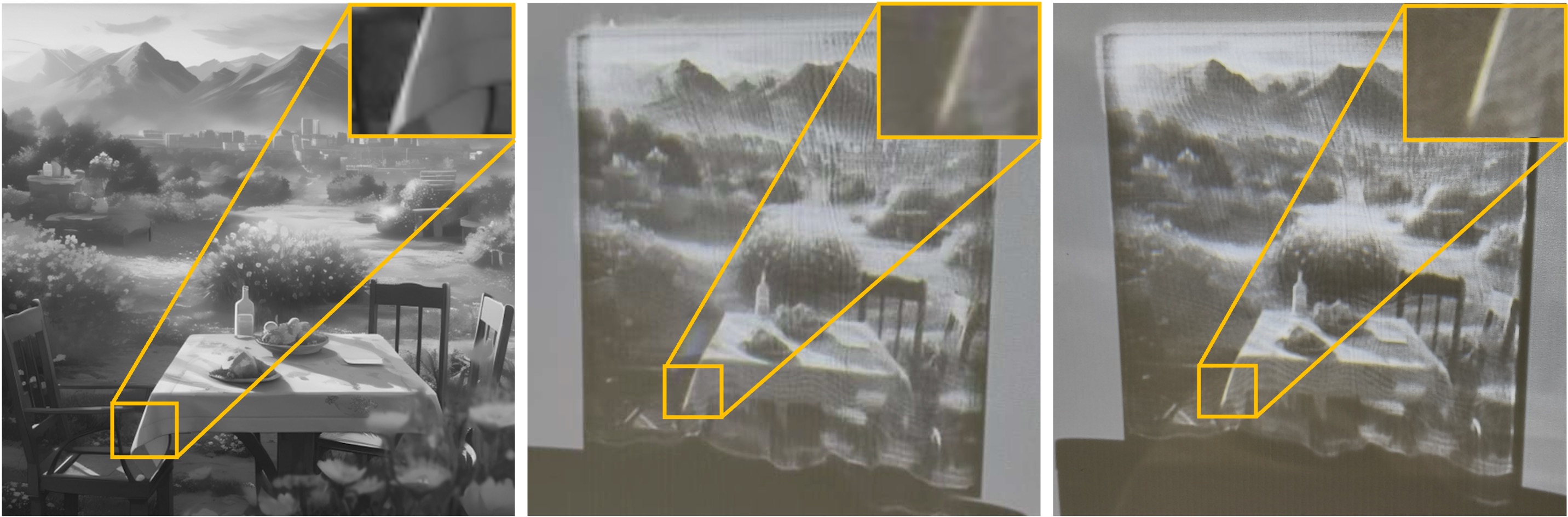}\\
  \vspace{-0.5em}
    \begin{picture}(0,0)
    \put(-108,-8){\footnotesize Target Image}
    \put(-31,-4){\footnotesize Point Light Source}
    \put(-8,-14){\footnotesize Based}
    \put(68,-8){\footnotesize Our Method}
  \end{picture}
  \vspace{0.8em}
  \caption{\label{physical_contrast}
  Comparison of physical results. From left to right: the target image, the caustic pattern from the lens designed using a point light source, and the pattern from the lens designed with our method.}
\end{figure}
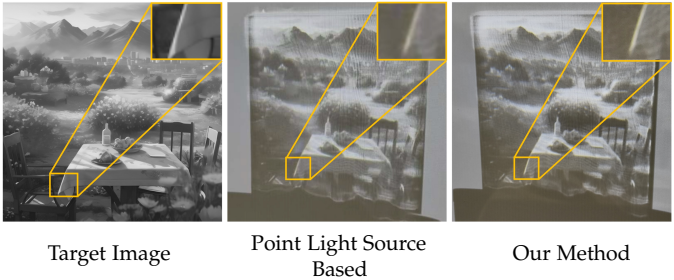

Figure~\ref{physical_contrast} presents the physical results from lenses designed using both a point light source and our proposed surface light source model. The manufacturing process inevitably introduces minor imperfections, which appear as stripes or ringing artifacts in the final image. These are inherent to the fabrication method (specifically the visible tool marks left by the ball-end mill during CNC machining), and consequently appear in both fabricated lenses. Despite these shared manufacturing limitations, a careful comparison reveals that the lens designed with our method produces a caustic image with details that more accurately match the target. This demonstrates the superior performance of our algorithm, as its advantages are evident even when accounting for real-world fabrication constraints.

\subsection{Ablation study}
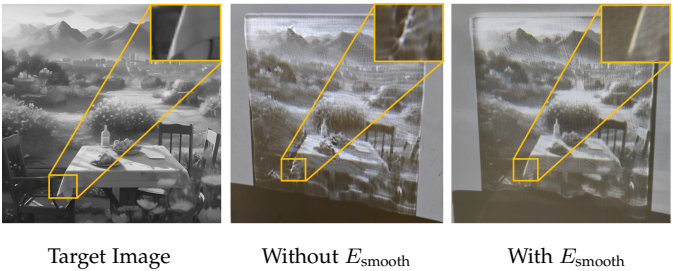
\begin{figure}[t]
  \centering
  \includegraphics[width=1.0\linewidth]{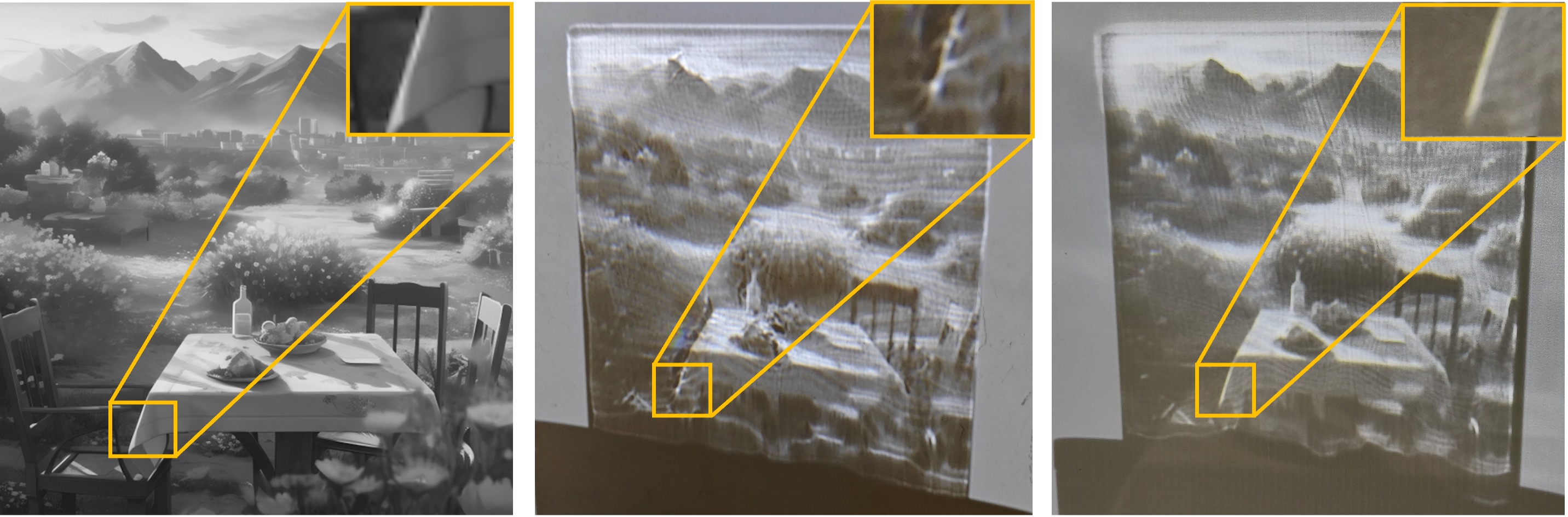}\\
  \vspace{-0.5em}
    \begin{picture}(0,0)
    \put(-108,-8){\footnotesize Target Image}
    \put(-27,-8){\footnotesize Without $E_{\text{smooth}}$}
    \put(66,-8){\footnotesize With $E_{\text{smooth}}$}
  \end{picture}
  \vspace{0.5em}
  \caption{\label{ablation_smoonth}
           Ablation study on the \(E_{\text{smooth}}\) loss term during the lens optimization process. From left to right: the target image, the physical experiment result without the \(E_{\text{smooth}}\) term, and the physical experiment result with the \(E_{\text{smooth}}\) term.}
\end{figure}

\begin{figure}[t]
  \centering
  \includegraphics[width=1.0\linewidth]{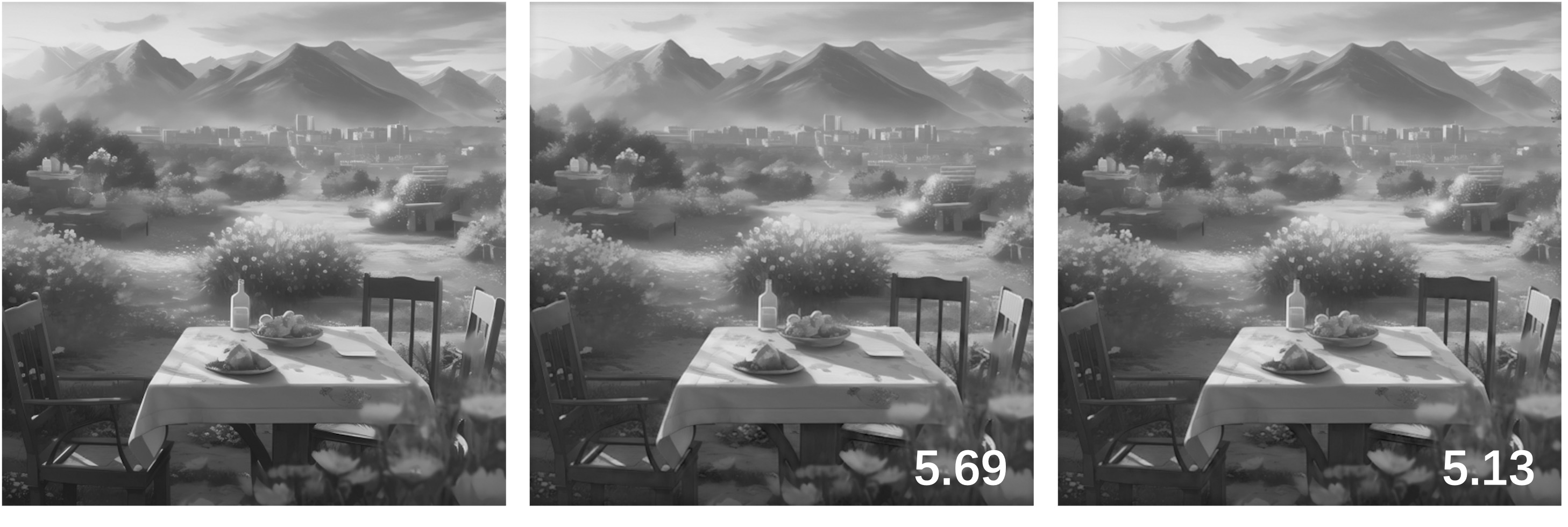}\\
  \vspace{-0.5em}
    \begin{picture}(0,0)
    \put(-108,-6){\footnotesize Target Image}
    \put(-36,-6){\footnotesize Without Contraction}
    \put(55,-6){\footnotesize With Contraction}
  \end{picture}
  \vspace{0.5em}
  \caption{\label{ablation_contraction}
       Ablation study on the contraction mapping. From left to right: the target image, rendering results of light source model optimized based on non contraction mapping, rendering results of light source model optimized based on contraction mapping.
    }
\end{figure}
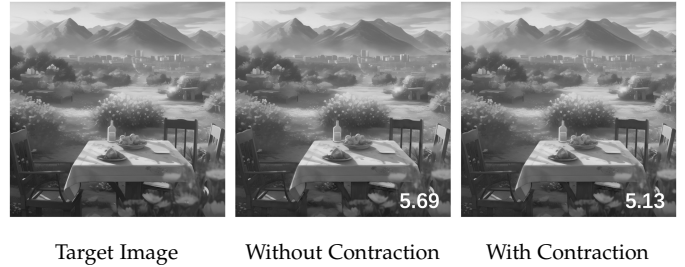

To verify the effectiveness of \(E_{\text{grad}}\), \(E_{\text{out}}\), and \(E_{\text{smooth}}\) in loss function~(\ref{optimization problem_caustic}), we conducted ablation studies on each of them individually.
Table~\ref{table3} presents the MAE and PSNR of the rendered results after removing \(E_{\text{grad}}\) or \(E_{\text{out}}\).
As shown in Table~\ref{table3}, the \(E_{\text{grad}}\) and \(E_{\text{out}}\) loss terms improve the quality of the rendering results to a certain extent.
Unlike the previous two terms, the term \(E_{\text{smooth}}\)  does not improve the evaluation metrics.
Instead, it results in a smoother lens surface after optimization, which reduces the difficulty of actual fabrication. 
We have included physical experiments for this term, as shown in Figure~\ref{ablation_smoonth}.
The introduction of the \(E_{\text{smooth}}\) term enables the fabricated lens to produce images that more closely match our designed results.
In addition, we conducted ablation experiments on whether to use contraction mapping during the light source optimization stage, as shown in Figure~\ref{ablation_contraction}. 
The experimental results indicate that contraction mapping can improve the accuracy of rendering results to a certain extent. 
More importantly, it limits the position of all optimized point light sources to the area of the real surface light source, making the light source model conform to physical reality.

\begin{table*}[t]
\centering
\caption{Partial results of the ablation experiment. We present the impact on experimental results when \(E_{\text{grad}}\), \(E_{\text{out}}\) are removed from the loss function~(\ref{optimization problem_caustic}).}
\small
\setlength{\tabcolsep}{8pt} 
\renewcommand{\arraystretch}{1.2} 

\begin{tabular}{l|cc|cc|cc}
\toprule 
\multirow{2}{*}{\textbf{Image}} 
& \multicolumn{2}{c|}{\textbf{Without $E_{grad}$}} 
& \multicolumn{2}{c|}{\textbf{Without $E_{out}$}} 
& \multicolumn{2}{c}{\textbf{With All}} \\
& MAE ($\times10^{-3}$) & PSNR (dB) & MAE ($\times10^{-3}$) & PSNR (dB) & MAE ($\times10^{-3}$) $\downarrow$ & PSNR (dB) $\uparrow$\\
\midrule
\textbf{Bridge} & 4.63 & 41.23 & 4.08 & 40.96 & 3.58 & 41.51 \\
\textbf{House} & 19.06 & 29.82 & 19.17 & 29.67 & 18.81 & 29.75 \\
\textbf{Flower} & 3.43 & 41.37 & 2.83 & 42.63 & 1.89 & 44.00 \\
\textbf{Table} & 6.45 & 38.27 & 5.76 & 38.23 & 5.13 & 38.78 \\
\textbf{Butterfly} & 3.72 & 41.93 & 3.80 & 41.43 & 3.07 & 42.66 \\
\textbf{Bear} & 2.25 & 46.32 & 2.18 & 46.01 & 1.93 & 47.12 \\
\bottomrule
\end{tabular}
\label{table3}
\end{table*}

\section{Conclusion and discussion}
In this paper, we propose a strategy to design the entire caustic process in two parts, incorporating the precise modeling of surface light sources into the lens shape design process.
For simulating illumination light sources, we propose a novel method for representing surface light sources and design a differentiable rendering framework, then utilize the differences between the caustic images of the input lenses and the reference image to drive the optimization of model parameters.
For the design of the lens shape, based on the optimized light source model, we guide the optimization of the free-form surface shape through the design of multiple types of penalty terms.
Experimental results show that our method better matches the illumination pattern of LED light sources, significantly improves the accuracy of caustic design, and achieves better detail performance.

Our method still has some limitations.
The light source model design in this paper is only for the most common planar LED light source. In future work, it can be considered to be extended to three-dimensional curved light sources to adapt to more lighting scenarios.

Some existing renderers, such as Mitsuba3, have implemented colored caustics at the simulation level.
In future work, our method could be extended to colored caustics by incorporating wavelength or color attributes into the parameters of each point light source and using different refractive indices for different colors or wavelengths during the lens refraction process.
Additionally, future work can incorporate CNC milling and polishing processes into the design of freeform surfaces to improve the precision of lens fabrication.

\section*{Acknowledgments}
This research was supported by the National Natural Science Foundation of China(No.62441224, No.62272433), the Fundamental Research Funds for the Central Universities, and the USTC Fellowship (No.S19582024).

\bibliographystyle{IEEEtran}
\bibliography{reference}
\begin{IEEEbiography}[{\includegraphics[width=1in]{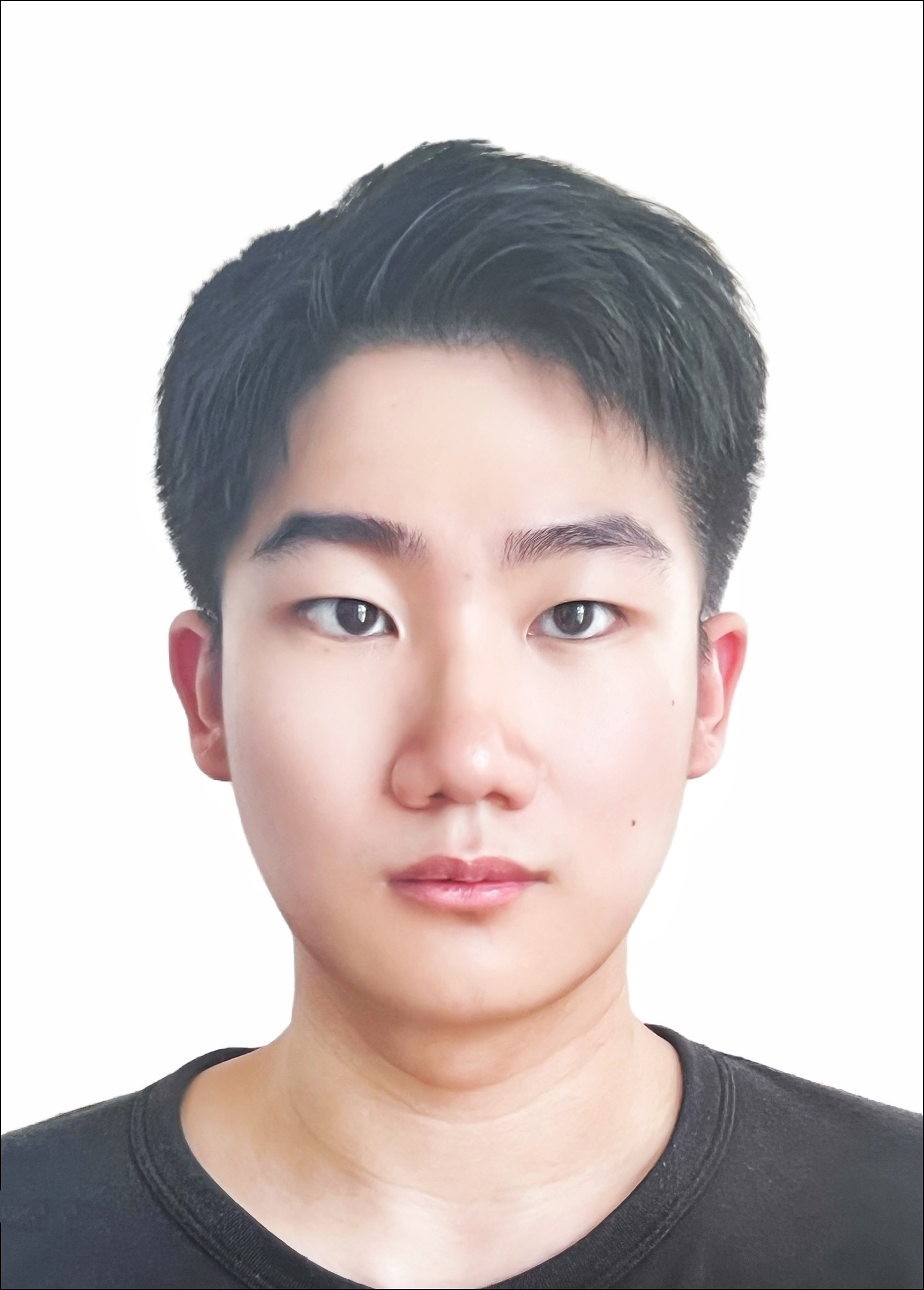}}]{Sizhuo Zhou} is currently working toward a PhD degree in the School of Mathematical Sciences, University of Science and Technology of China. His research interests include computer graphics and non-image optics. In addition, he is also interested in autonomous driving and 3D AIGC.
\end{IEEEbiography}
\begin{IEEEbiography}[{\includegraphics[width=1in]{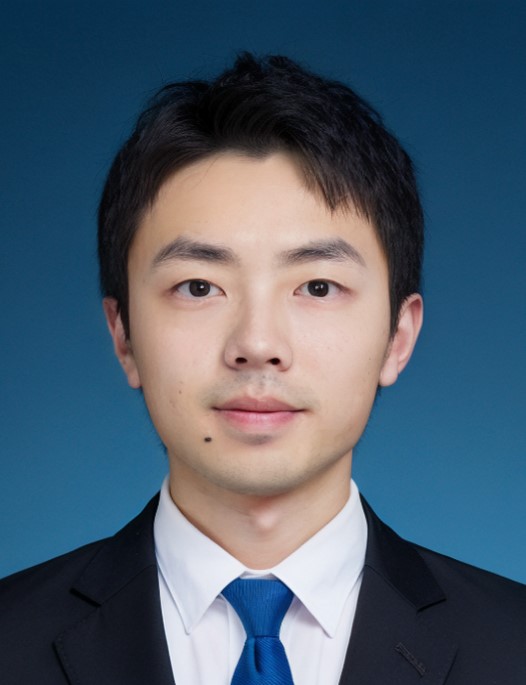}}]{Yuou Sun} is currently working toward a PhD degree in the School of Mathematical Sciences, University of Science and Technology of China. His research interests include computer graphics, numerical optimization and non-image optics.
\end{IEEEbiography}
\begin{IEEEbiography}[{\includegraphics[width=1in]{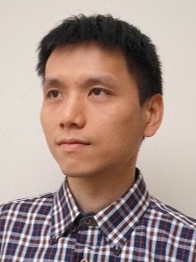}}]{Bailin Deng}
is a senior lecturer in the School of Computer Science and Informatics at Cardiff University. He received the BEng degree in computer software (2005) and the MSc degree in computer science (2008) from Tsinghua University (China), and the PhD degree in technical mathematics (2011) from Vienna University of Technology (Austria). His research interests include geometry processing, numerical optimization, computational design, and digital fabrication. He is a member of the IEEE.
\end{IEEEbiography}

\begin{IEEEbiography}[{\includegraphics[width=1in]{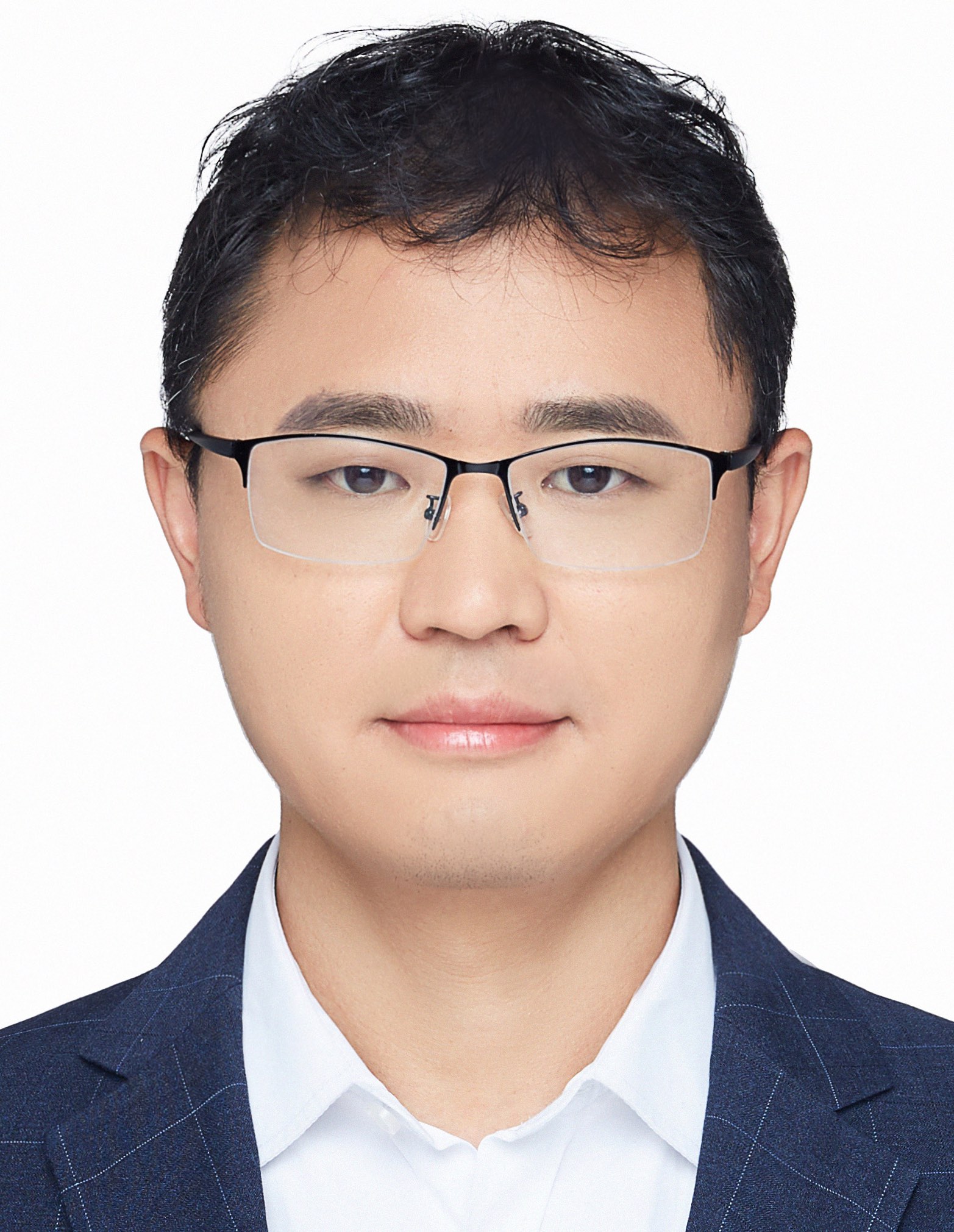}}]{Juyong Zhang}
is a professor in the School of Mathematical Sciences at University of Science and Technology of China. He received the BS degree from the University of Science and Technology of China in 2006, and the Ph.D degree from Nanyang Technological University, Singapore. He mainly conducts research at the intersection of Vision, Graphics, and AI with a special focus on capturing, modeling and synthesizing objects, humans and large-scale scenes. He is an associate editor of IEEE Transactions on Mobile Computing and IEEE Computer Graphics and Applications.
\end{IEEEbiography}

\end{document}